\documentclass[ALICE,manyauthors]{cernphprep}
\usepackage[comma,square,numbers,sort&compress]{natbib}
\RequirePackage{lineno}

\newcommand{\Ls}{\rm{\Lambda^{*}}}
\newcommand{\Lam}{\rm{\Lambda}}
\newcommand{\sn}{\sqrt{s_{\rm{NN}}}}
\newcommand{\s}{\sqrt{s}}
\newcommand{\pt}{p_{\rm{T}}}
\newcommand{\mpt}{\langle p_{\rm{T}}\rangle}
\newcommand{\dnchdeta}{\langle {\rm d}N_{\rm ch}/{\rm d} \eta_{\rm lab} \rangle}

\usepackage[usenames,dvipsnames,svgnames,table]{xcolor}

\usepackage{rotating}
\usepackage{graphicx}
\usepackage{amssymb}
\usepackage{epstopdf}
\usepackage{subfigure}
\usepackage{caption}
\usepackage{lipsum}
\usepackage{hyperref}
\usepackage[T1]{fontenc}
\begin{document}

%
%
\begin{titlepage}
\PHyear{2019}       
\PHnumber{178}      
\PHdate{19 August}  

%
%
\title{Measurement of $\Lam$(1520) production in pp collisions at $\s$ $\mathrm{=}$ 7 TeV and p--Pb collisions at $\sn$ $\mathrm{=}$ 5.02 TeV}
\ShortTitle{Measurement of $\Lam$(1520) production in pp and p--Pb collisions} 
\author{ALICE Collaboration}
%
\ShortAuthor{}  

\begin{abstract}
The production of the $\Lam$(1520) baryonic resonance has been measured at midrapidity in inelastic pp collisions at $\s$ $\mathrm{=}$ 7 TeV and in p--Pb collisions at $\sn$ $\mathrm{=}$ 5.02 TeV for non-single diffractive events and in multiplicity classes. The resonance is reconstructed through its hadronic decay channel $\Lam$(1520) $\rightarrow$ pK$^{-}$ and the charge conjugate with the ALICE detector. The integrated yields and mean  transverse momenta are calculated from the measured transverse momentum distributions in pp and p--Pb collisions. The mean transverse momenta follow
 mass ordering as previously observed for other hyperons in the same collision systems.
A Blast-Wave function constrained by other light hadrons ($\pi$, K, K$_{\rm{S}}^0$, p, $\Lam$) describes the shape of the $\Lam$(1520) transverse momentum distribution up to 3.5 GeV/$c$ in p--Pb collisions.
In the framework of this model, this observation suggests that the $\Lam$(1520) resonance participates in the same collective radial flow as other light hadrons.
The ratio of the yield of $\Lam(1520)$ to the yield of the ground state particle $\Lam$
 remains constant as a function of charged-particle multiplicity, suggesting that there is no net effect of the hadronic phase in p--Pb collisions on the $\Lam$(1520) yield.
\end{abstract}
\end{titlepage}
%
%
%
%

\setcounter{page}{2} 

\section{Introduction} \label{sec:Introduction}
High-energy heavy-ion (A--A) collisions offer a unique possibility to study  nuclear matter under extreme temperature and density, in particular the properties of the deconfined quark--gluon plasma (QGP)~\cite{Adams:2005dq, Adcox:2004mh, Arsene:2004fa, Back:2004je, Aamodt:2010pa, Aamodt:2010jd, Schukraft:2011na}, which was predicted by quantum chromodynamics (QCD)~\cite{Cabibbo:1975ig, Shuryak:1978ij, Laermann:2003cv, Aoki:2006we, Gupta:2011wh}. The interpretation of the heavy-ion results depends crucially on the comparison with results from small collision systems such as proton--proton (pp) or proton--nucleus (p--A). 
Measurements in pp collisions establish a reference for larger systems and are used to test perturbative QCD models. The p--A collisions, which are intermediate between pp and A--A collisions in terms of system size and number of produced particles~\cite{ALICE:2012xs,  Adam:2014qja, Aamodt:2010cz}, are traditionally used to separate initial and final-state effects~\cite{Arneodo:1992wf, Abelev:2014hha}. 
However, at the LHC the pseudorapidity density (d$N_{\rm ch}$/d$\eta$) of final-state charged particles in pp and p--A collisions can reach values comparable to those achieved in semi-peripheral Au--Au~\cite{Alver:2010ck} and Pb--Pb collisions~\cite{ALICE:2017jyt} at the top energies of RHIC and the LHC, respectively.
 Therefore, there exists a possibility of final-state effects due to the formation of dense matter even in p--A collisions. 
 
During the evolution of the systems formed in A--A or p--A collisions, the yields of short-lived resonances may be influenced by interactions in the late hadronic phase.
The re-scattering of the decay products in the medium may prevent the detection of a fraction of the resonances, whereas pseudo-elastic hadron scattering can regenerate them. The strengths of the re-scattering and regeneration effects depend on the scattering 
cross sections of the decay products, the particle density of the produced medium, the lifetimes of the resonances and the lifetime of the hadronic phase.
The latter can be studied by comparing yields of short-lived resonances with different lifetimes to yields of long-lived particles~\cite{Aggarwal:2010mt, Abelev:2014uua, Adam:2017zbf, Adams:2006yu}. ALICE has observed that in the most central p--Pb and Pb--Pb collisions~\cite{Abelev:2014uua, Adam:2017zbf, Adam:2016bpr} the K$^{*0}$/K ratio is significantly suppressed with respect to peripheral collisions, pp collisions, and predictions of statistical hadronization models~\cite{Stachel:2013zma, Vovchenko:2019kes}. 
A similar suppression is also observed for $\rho^{0}/\pi$ ratio in central Pb--Pb collisions with respect to peripheral Pb--Pb collisions, pp collisions, and predictions of statistical hadronization models~\cite{Acharya:2018qnp}.
No suppression is observed for the ${\rm \phi}$/K ratio,  as the ${\rm \phi}$ meson lives ten times longer than the K$^{*0}$.
To provide more insight into the properties of the hadronic phase, other resonances whose lifetimes are in between those of the K$^{*0}$ ($\tau_{\rm{K}^{*0}}$ $\mathrm{=}$ 4.17 $\pm$ 0.04 fm/$c$~\cite{Tanabashi:2018oca}) and $\phi$ ($\tau_{\rm \phi}$ $\mathrm{=}$ 46.4 $\pm$ 0.14 fm/$c$~\cite{Tanabashi:2018oca}) should be studied.
The $\Lam(1520)$ resonance 
is a strongly decaying particle having $\tau_{\Lam(1520)}$ $\mathrm{=}$ 12.6 $\pm$ 0.8 fm/$c$~\cite{Tanabashi:2018oca}. This makes the study of the $\Lam(1520)$ resonance important for understanding the evolution of the system.
Previously, the STAR experiment at RHIC measured $\Lam(1520)$ production in pp, d--Au and Au--Au collisions at a center-of-mass energy per nucleon pair ($\sn$) of 200 GeV \cite{Adams:2006yu, Abelev:2008yz} and showed a hint of suppression of the $\Lam(1520)/\Lam$ yield ratio in central Au--Au collisions compared to the values observed in pp and d--Au collisions.  
A measurement of the $\Lam(1520)$ in Pb--Pb collisions at $\sn$ $\mathrm{=}$ 2.76 TeV was reported in~\cite{ALICE:2018ewo}. The $\Lam(1520)/\Lam$ yield ratio is found to be suppressed in central (0--20\%) Pb--Pb collisions relative to peripheral (50--80\%) Pb--Pb collisions. The suppression factor is found to be 0.54$\pm$0.08(stat)$\pm$0.12(sys). The EPOS3~\cite{Drescher:2000ha, Werner:2010aa, Werner:2013tya} event generator, which incorporates the UrQMD model~\cite{Bass:1998ca} to simulate the hadronic phase, predicts a significant suppression of the $\Lam(1520)/\Lam$ yield ratio in central Pb--Pb collisions at $\sn$ $\mathrm{=}$ 2.76 TeV~\cite{Knospe:2015nva}. However, the corresponding Pb--Pb measurements show a stronger suppression than predicted by EPOS3. This motivates the study of the $\Lam(1520)$ resonance in different collision systems at the LHC in order to better understand the properties of the hadronic phase. The pp and p--Pb data studied in this paper thus provide important baseline measurements for the corresponding results in Pb--Pb collisions.

In addition, several measurements~\cite{Abelev:2012ola, Abelev:2013wsa, Abelev:2014mda} in p--A collisions indicate that these systems cannot be explained as an incoherent superposition of pp collisions, rather suggesting~\cite{Bozek:2011if, Bozek:2012gr} the presence of collective effects.
In p--Pb collisions, a significant increase of the average transverse momentum as a function of charged particle density has been observed~\cite{Abelev:2013haa} and this is reminiscent of the effect observed in Pb--Pb collisions, where it is interpreted as a consequence of radial flow. The measurement of the $\pt$ spectra of  the $\Lam$(1520) resonance can further confirm such effects, and thus can be used to better constrain the properties of the collective radial expansion.

Recently, the ALICE Collaboration reported measurements of multi-strange particles in p--Pb collisions~\cite{Adam:2015vsf}. 
The hyperon-to-pion ratios increase with multiplicity in p--Pb collisions, and range from the values measured in pp to the those in Pb--Pb collisions. The rate of the increase is more pronounced for particles with higher strangeness content. 
Therefore, it will be interesting to study the production of excited strange hadrons, like $\Lam$(1520), $\Xi$(1530), as a function of multiplicity. Doing so in p--Pb collisions would help bridge the gap between the pp and Pb--Pb collision systems. 

Throughout this paper, the $\Lam$(1520) resonance will be referred as $\Ls$. The invariant mass of $\Ls$ is reconstructed through its hadronic decay channel $\Ls\rightarrow$ pK$^{-}$, with a  branching ratio of BR $\mathrm{=}$ (22.5 $\pm$ 1)\%~\cite{Tanabashi:2018oca}.
 The invariant mass distributions of the pK$^-$ and $\overline{\rm p}$K$^+$ were combined to reduce the statistical uncertainties. Therefore in this paper, unless specified, $\Ls$ denotes $\Ls$+$\overline{\Lam}^*$. 
The paper is organized as follows. The experimental setup is briefly presented in Section~\ref{sec:ExperimentalSetup}. 
Section~\ref{sec:DataSample} describes the data samples and event selection. 
Section~\ref{sec:AnalysisDetails} illustrates the analysis procedure as well the determination of the systematic uncertainties. The results are discussed in Section~\ref{sec:ResultsandDiscussion}, and a summary is provided in Section~\ref{sec:SummaryandConclusion}.
\section{Experimental setup} \label{sec:ExperimentalSetup} 
The ALICE~\cite{Aamodt:2008zz, Abelev:2014ffa} detector is specifically designed to study a variety of observables in the high-multiplicity environment achieved in central A--A collisions at LHC energies.
The detector is optimized to reconstruct and identify particles produced in the collisions over a wide momentum range.

In this analysis, only the central barrel sub-detectors were used for track reconstruction. These detectors have a common pseudorapidity coverage in the laboratory frame of $|\eta_{\rm{lab}}| < 0.9$, and are placed in a solenoidal 0.5 T magnetic field directed along the beam axis. 
The Inner Tracking System (ITS)~\cite{Aamodt:2010aa} provides high resolution tracking points close to the beam line. 
The ITS is composed of six cylindrical layers of silicon detectors, located at radial distances between 3.9 and 43 cm from the beam axis. The two innermost layers are Silicon Pixel Detectors (SPD), the two intermediate ones are Silicon Drift Detectors (SDD), and the two outermost ones are Silicon Strip Detectors (SSD). 
The Time Projection Chamber (TPC)~\cite{Alme:2010ke} is the main tracking detector of the central barrel. The TPC is a cylindrical drift chamber, and covers the radial distance $85 < r < 247$ cm. In addition to tracking, the TPC is used for the identification of particles via their specific ionization energy loss  d$E$$\mathrm{/}$d$x$ as they pass through the active gas region of the TPC.  
The separation power of particle identification in the TPC defined in terms of standard deviations as a function of particle momentum is discussed in~\cite{Abelev:2014ffa}.
This analysis uses charged tracks, which are reconstructed using tracking information, both in the ITS and in the TPC. 
The Time-Of-Flight (TOF) detector~\cite{Akindinov:2013tea} is an array of Multi-gap Resistive Plate Chambers (MRPC). The time resolution of TOF is about 85 ps, increasing to about 120 ps due to a worse start-time (collision-time) resolution in the case of low multiplicity events~\cite{Abelev:2014ffa}.
The TOF is located at a radial distance of  $370 < r < 399$ cm from the beam axis.
The purpose of this detector is to identify particles using the time-of-flight, together with the momentum and path length measured with the ITS and the TPC.
 The TOF can separate  pions from kaons and protons by twice its resolution, for momenta up to 2.5 and 4 GeV/$c$, respectively.

Forward detectors, such as the V0, T0, and Zero-Degree Calorimeters (ZDC)~\cite{Abbas:2013taa, Cortese:781854, Gallio:381433}, are used for triggering and event characterization.
The V0 consists of two arrays of 32 scintillator detectors. They cover the full azimuthal angle in the pseudorapidity regions $2.8 < \eta_{\rm{lab}} < 5.1$ (V0A) and $-3.7 < \eta_{\rm{lab}} < -1.7$ (V0C). 
 The V0 can be used to define the event multiplicity interval. The T0 consists of two arrays of quartz Cherenkov counters, T0A ($4.6 < \eta < 4.9$) and T0C ($ -3.3 < \eta < -3.0$), and provides the time and the longitudinal position of the interaction. 
The ZDC is a hadronic calorimeter consisting of two W-quartz neutron and two brass-quartz proton calorimeters, placed symmetrically at a distance of 113 m on both sides of the interaction point. In this analysis the ZDC is used for background rejection.
%
%
\section{Data sample and event selection} \label{sec:DataSample} 
The data samples analysed in this paper were recorded during the LHC pp run in 2010, and the p--Pb run in 2013. For pp collisions, the center-of-mass 
energy is 7 TeV and the analysis is carried out within the rapidity range $-0.5 < y < 0.5$. 
 The instantaneous luminosity at the ALICE interaction point was in the range 0.6--1.2$\times$10$^{29}$ cm$^{-2}$s$^{-1}$. 
This limited the collision pile-up probability to an average rate of 2.5\%~\cite{Adam:2015qaa}.
 For p--Pb collisions at $\sn$ $\mathrm{=}$ 5.02 TeV, the beam energies are 4 TeV for protons, and 1.577 TeV for Pb. For the analyzed data set, the Pb beam was circulated towards the positive rapidity direction (labelled as ALICE ``A'' side); conversely, the proton beam was circulated towards negative rapidity direction (labelled as ALICE ``C'' side). 
The asymmetry in the energies of the proton and Pb beam shifts the nucleon--nucleon centre-of-mass system, relative to the laboratory frame, by 0.465 units of rapidity along the proton beam direction.
In the following, the variables $y_{\rm{lab}}$ ($\eta_{\rm{lab}}$) are used to indicate the rapidity (pseudorapidity) in the laboratory reference frame, whereas $y \ (\eta)$ denotes the same in the center-of-mass frame. The analysis presented in this paper was performed in the rapidity window $-0.5 < y < 0$. The peak luminosity during data taking was about 10$^{29}$ cm$^{-2}$s$^{-1}$ with a probability of multiple interactions below 3\%~\cite{Abelev:2014oea}. 
The small fraction of pile-up events from the same bunch crossing was removed by rejecting events with multiple vertices using the SPD.
 Pile-up of collisions from different bunch crossings is negligible due to the bunch-crossing spacing (200 ns) being larger than the integration time of the ZDC. 

  The pp collision data were collected using a minimum-bias (MB) trigger. This trigger required a signal in the SPD or in any one of the V0 scintillator arrays in coincidence with a bunch crossing. 
With this configuration about 85\% of all inelastic events were triggered~\cite{Adam:2015gka}. The V0 detector measures the event time with a resolution of about 1 ns. Using the timing information from the V0 detector, the contamination due to beam-induced background is removed offline. 
Selected events are further required to have a reconstructed primary vertex within $\pm$ 10 cm from the center of the ALICE detector, along the beam axis to ensure a symmetric rapidity coverage of the barrel detectors, and to reduce the remaining beam-gas contamination.
The data analysis is carried out using a sample of $\sim$125 million minimum-bias pp collisions.   
   
In the p--Pb data sample, the events were selected using the trigger condition requiring a logical AND between signals in V0A and V0C. 
This reduces the contamination from single-diffractive and electromagnetic interactions. These non-single-diffractive (NSD) events include double-diffractive interactions, where both colliding nucleons break up by producing particles separated by a large rapidity gap. The trigger and event selection efficiency for NSD events is $\epsilon_{NSD}$ $\mathrm{=}$ 99.2\% as described in Ref.~\cite{ALICE:2012xs, Adam:2014qja}. 
The beam induced background was further reduced offline using timing cuts on the signals from the V0 and ZDC detectors~\cite{Abelev:2014ffa}. The same procedure as in pp collisions was used to reconstruct the primary vertex. The event sample amounts to a total of about 100 million accepted events.  The NSD events were further divided into four multiplicity intervals according to the charge deposited in the forward V0A detector, positioned along the direction of the Pb beam~\cite{Adam:2014qja}.  The multiplicity intervals and their corresponding mean charged-particle density ($\langle  {\rm d}N_{\rm{ch}}/{\rm d}\eta_{\rm{lab}} \rangle$) measured at midrapidity ($|\it{\eta}_{\rm{lab}}| <$ 0.5) are given in Table~\ref{table:chargemult}.

\begin{table}[h]
\centering
\caption{Average charged-particle multiplicity density measured at midrapidity in the used event multiplicity intervals in p--Pb collisions at $\sn$ $\mathrm{=}$ 5.02 TeV~\cite{Adam:2014qja} and in inelastic pp collisions at \mbox{$\s$ $\mathrm{=}$ 7 TeV~\cite{Adam:2015gka}}.}
\begin{tabular*}{\columnwidth}{@{\extracolsep{\fill}}ccc@{}}
\hline
System & Event Class  &$\langle  {\rm d}N_{\rm{ch}}/{\rm d}\eta_{\rm{lab}} \rangle_{|\eta_{\rm{lab}}| < 0.5} $ \\
\hline
\hline
p--Pb & 0--20\%    &  35.6 $\pm$ 0.8 \\
& 20--40\%  & 23.2 $\pm$ 0.5  \\
& 40--60\%  & 16.1 $\pm$ 0.4 \\
& 60--100\% & 7.1 $\pm$ 0.2 \\
& 0--100\%  & 17.4 $\pm$ 0.7\\
\hline
pp & INEL & 4.6$^{+0.3}_{-0.2}$   \\
\hline
\hline
\end{tabular*}
\label{table:chargemult}
\end{table}

%
%
%
\section{Analysis details} \label{sec:AnalysisDetails} 
 
\subsection{Track cuts and particle identification}

This analysis uses tracks reconstructed in the TPC and ITS. Each track is required to have at least one hit in one of the two layers of the SPD. 
The criteria for selecting a reconstructed track in the TPC are the following: the track is required to cross at least 70, out of the maximum 159, horizontal readout segments (or ``rows'') along the transverse plane of the TPC; and the ratio of crossed rows over findable clusters in the TPC has to be greater than 0.8. In addition to this, standard ALICE track quality cuts have been applied~\cite{Adam:2016bpr}. These selections limit the contamination from secondary and fake tracks, while ensuring a high efficiency, and good d$E$/d$x$ resolution. 
Tracks with transverse momentum $\pt < 0.15$ GeV/$c$ and $|\eta_{\rm{lab}}| > 0.8$ are rejected to suppress boundary effects.
Due to the small lifetime of $\Ls$, the daughter particles should be reconstructed as primary tracks originating from the event vertex. For this purpose, we used distance of closest approach (DCA) cuts, along the transverse and $z$ directions of 7$\sigma_{\rm{DCA}}$($\pt$) and 2 cm, respectively. Here, $\sigma_{\rm{DCA}}$ is $\pt$-dependent and parameterized  as 0.0015 $+$ 0.005/$\pt^{1.1}$~\cite{Abelev:2014ffa},  where $\pt$ is measured in units of GeV/$c$.
The methods used for charged particle identification (PID) in pp and p--Pb collisions are as follows. 
In pp collisions, low momentum protons ($p <$ 1.1 GeV/$c$) and \mbox{K ($p <$ 0.6 GeV/$c$)} are selected using the TPC with a 3$\sigma_{\rm TPC}$ PID cut on the measured d$E$$\rm{/}$d$x$ distribution. 
Higher momentum tracks are identified by requiring that the measured time-of-flight and d$E$$\rm{/}$d$x$ do not deviate from their expected values for each given mass hypothesis by more than 3$\sigma_{\rm TOF}$ and 5$\sigma_{\rm TPC}$, respectively (see~\cite{Adam:2015qaa, Aamodt:2011zj} for a discussion of the particle identification using TPC and TOF). 
For p--Pb collisions, tracks of any momentum that have a hit in the active TOF region are identified with a 3$\sigma_{\rm TOF}$ and 5$\sigma_{\rm TPC}$ cut, on the measured time-of-flight and d$E$$\rm{/}$d$x$ 
 values. 
If there is no hit in the active TOF region, the d$E$$\rm{/}$d$x$ of low momentum (defined earlier) tracks are required to be within 3$\sigma_{\rm TPC}$ of their expected values for each given mass hypothesis. This PID selection reduces the misidentification of particles over a large momentum region, hence reducing the combinatorial background under the signal peak. The rapidity cuts for pK pairs in pp and p--Pb collisions are $|y| < $ 0.5 and $-0.5 < y <$ 0, respectively. 

\subsection{Invariant mass reconstruction}
\begin{figure}[h] 
\centering
\includegraphics[scale=0.75]{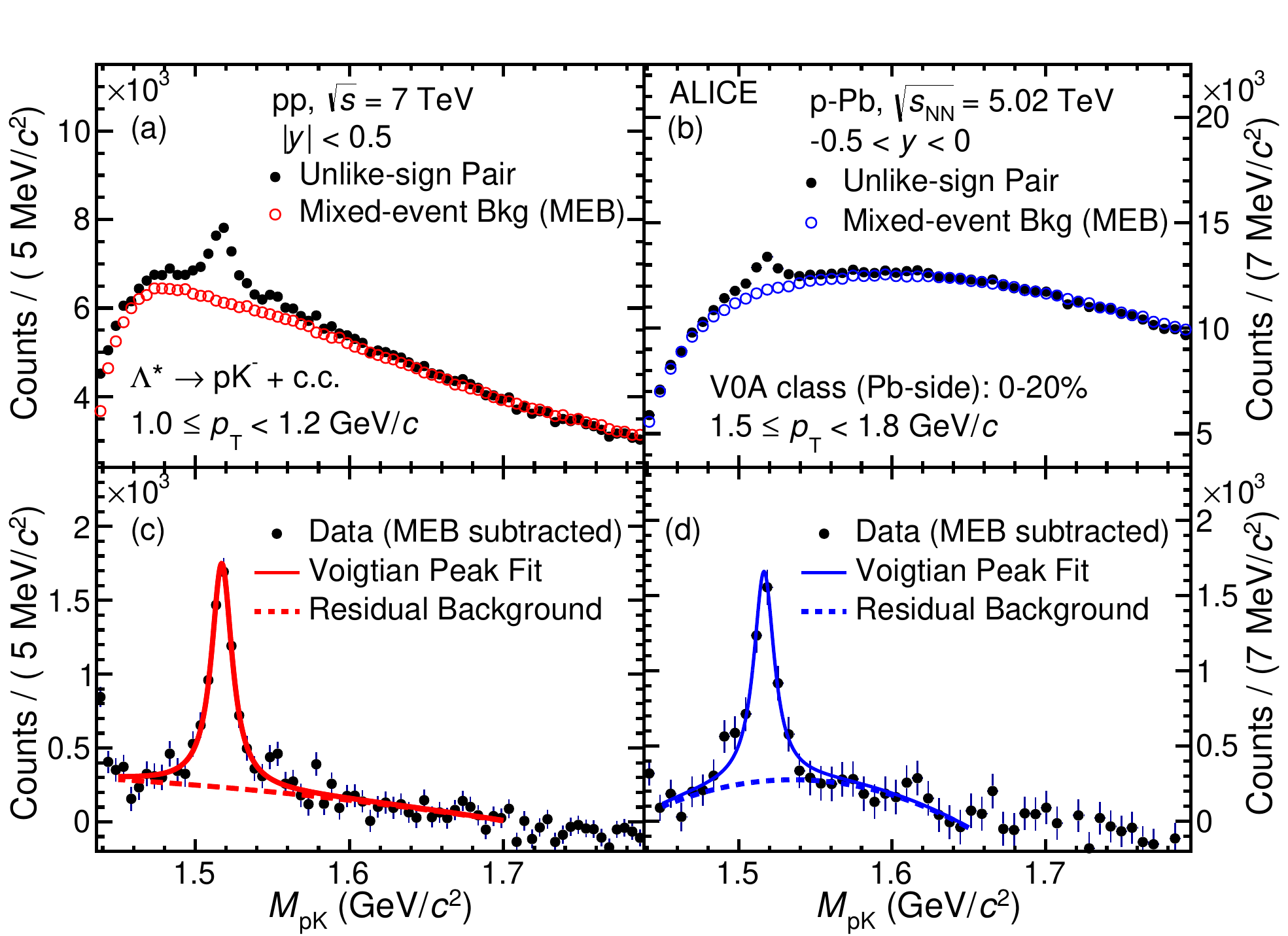} 
\caption{Invariant-mass distributions of pK pairs for MB pp collisions at $\sqrt{s}$ $=$ 7 TeV (left panels) and for p--Pb collisions in the 0--20\% multiplicity interval at $\sn$ $\mathrm{=}$ 5.02 TeV (right panels) for the momentum interval 1.0 $\leq \pt < $ 1.2 GeV/$c$ and 1.5 $\leq \pt < $ 1.8 GeV/$c$, respectively. Panels (a) and (b) show the unlike-sign pK invariant-mass distribution from the same event and normalized combinatorial background for pp and p--Pb collisions, respectively. Panels (c) and (d) show the invariant-mass distribution after subtraction of the  combinatorial background. The solid curve represents the Voigtian fit, while the dashed line describes the residual background. The statistical uncertainties are shown as bars.}
\label{fig:signalpppPb}
\end{figure}

Invariant mass distributions are reconstructed by combining pairs of oppositely charged pK pairs in the same event.
Examples of pK invariant-mass distributions are presented in the panels (a) and (b) of Fig.~\ref{fig:signalpppPb} for pp and p--Pb collisions, respectively. 
Clear peaks of $\Ls$ can be observed in the figures, which sit on top of combinatorial backgrounds.
The uncorrelated background is estimated using the mixed-event technique (ME). However at low momentum, $\pt <$ 1 GeV/$c$ ( 1.2 GeV/$c$) for pp (p--Pb) collisions, the like-sign technique (LS) was used since it better described the background shape.
In the ME approach, each proton track in an event was combined with kaon tracks from 10 different events. In order to minimize distortions, and to ensure a similar event structure, events with vertex position differences within 1 cm in the $z$ direction and charged-particle multiplicity differences within 10 have been chosen for mixing. For the LS method, the background is constructed by combining like-charged pairs of pK (pK$^+$ and $\overline{\rm p}$K$^-$) from the same event to get the geometrical mean of the two distributions,  2$\sqrt{\rm{N_{pK^+} \times N_{\overline{\rm p}K^-}}}$ in each invariant mass bin.
These background distributions are normalized in the mass region 1.7 $<$ $M_{\rm{pK}}$ $<$ 1.85 GeV/$c^{2}$ well outside the signal peak.
The background-subtracted distributions are shown in the panels (c) and (d) of Fig.~\ref{fig:signalpppPb} for MB pp collisions, and the 0--20\% multiplicity interval of p--Pb collisions, respectively.  These distributions exhibit characteristic peaks on top of residual backgrounds. These leftover backgrounds are mainly due to correlated pairs from jets, multi-body decays of heavier particles or correlated pairs from real resonance decays, misidentified as p or K during PID selection. A study of Monte Carlo simulations was performed to ensure that the shape of the correlated background is a smooth function of mass which 
can be well described with a second order polynomial in the fitting range.
Each signal peak has been fitted with a Voigtian function (convolution of a Breit-Wigner and a Gaussian) on top of a second order polynomial in invariant mass as in~\cite{Abelev:2012hy}. This is shown in the lower panels of Fig.~\ref{fig:signalpppPb}.
The polynomial is used to describe the residual background and the Voigtian function gives the resonance mass, width and yield.
The Gaussian component of the Voigtian function accounts for the mass resolution of the experimental setup.
A study of Monte Carlo simulations was done to estimate this mass resolution. 
\\
The mass resolution was found to depend on $\pt$. At low $\pt$ it shows a decreasing trend and reaches its lowest value of 1 MeV/$c^2$ at $\pt$ $\mathrm{=}$ 1 GeV/$c$,  then monotonically increases to a value of 1.6 MeV/$c^2$ at $\pt$ $\mathrm{=}$ 6 GeV/$c$. It is important to note that the mass resolution has very limited effect on the reconstructed peak shape due to large intrinsic width of the $\Ls$ resonance ($\Gamma$ $\mathrm{=}$ 15.73 MeV/$c^2$)~\cite{Tanabashi:2018oca}. The raw yield is calculated by integrating the invariant mass distribution after the subtraction of the combinatorial background and subtracting the integral of the residual background function in the range ($M-2\Gamma$, $M+2\Gamma$), where  $M$ is the mass peak position that comes from the Voigtian fit and $\Gamma$ is the PDG accepted value for the width of $\Ls$. The fractions of the yields in the tails on both sides of the peak are calculated from the fit function and applied as corrections to the measured yields.

\subsection{Detector acceptance and efficiency}
\begin{figure}[h] 
\begin{minipage}{\columnwidth}
\centering
\includegraphics[width=0.5\textwidth]{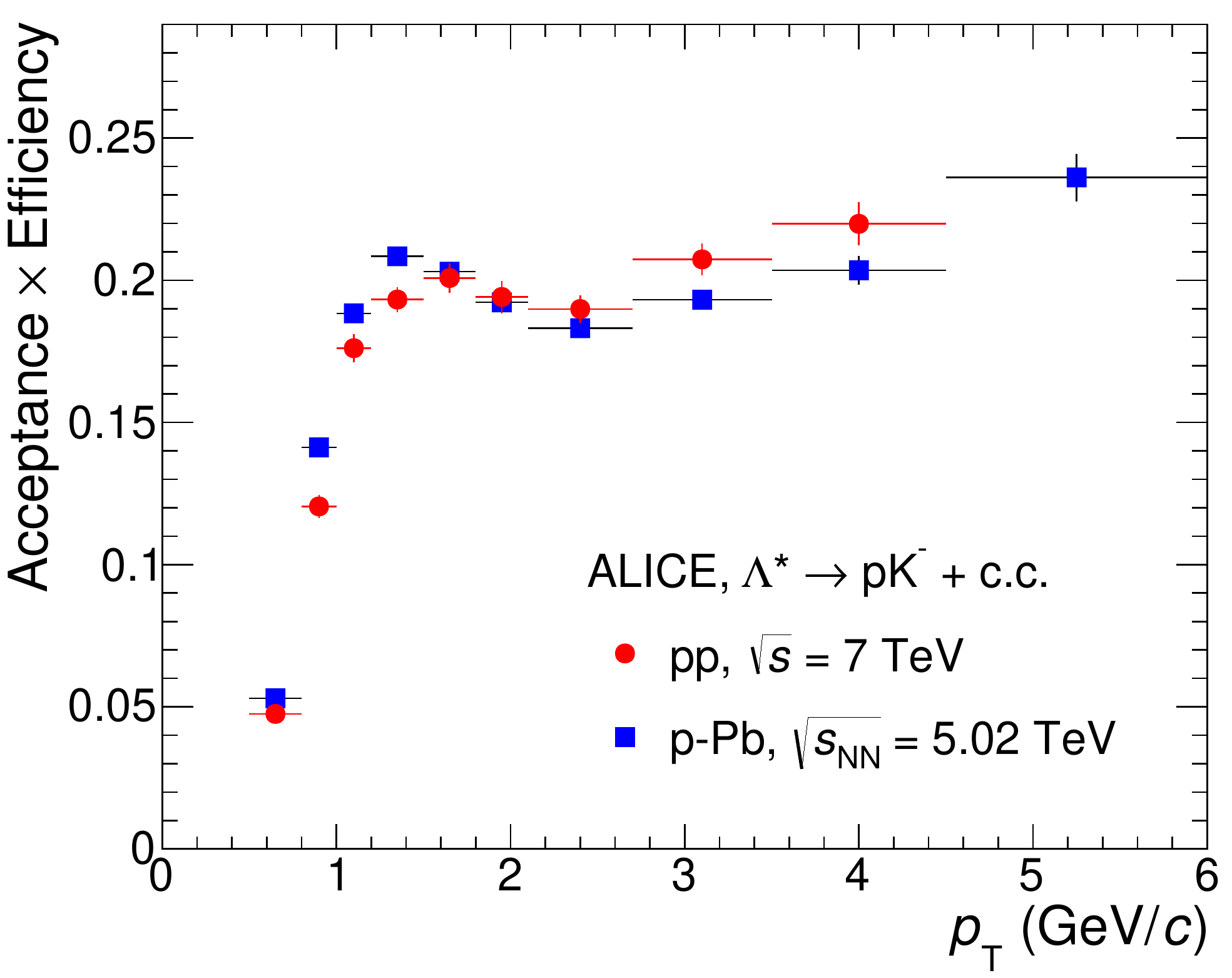} 
\end{minipage}
\caption{The geometrical acceptances times reconstruction efficiency (A $\times \  \epsilon$) for $\Ls$ in minimum bias pp and p--Pb events. Uncertainties (bars) are statistical only.}
\label{fig:efficiency}
\end{figure}

In order to evaluate the detector acceptance and reconstruction efficiency ($A \times \epsilon$), Monte Carlo pp and \mbox{p--Pb} events were simulated using the PYTHIA Perugia 2011~\cite{Skands:2010ak} and DPMJET 3.05~\cite{Roesler:2000he} event generators, respectively. The detector geometry and material budget were modeled by GEANT3~\cite{Brun:1994aa}, which is also used for the propagation of particles through detector material.
   The acceptance and efficiency corrections were determined as the fractions of generated resonances that were reconstructed in the rapidity interval $|y| < 0.5$ for pp events and  $-0.5 < y <$ 0 for p--Pb events. The selected primary p and K tracks have to pass the same kinematic, track selection, and PID cuts as applied in the real data. 
Since the generated $\Ls$ $\pt$ spectrum has a different shape than the measured $\pt$ spectrum, it is therefore necessary to weight the generated $\pt$ spectrum so that it has the shape of the measured spectrum. The $A \times \epsilon$ obtained after applying this re-weighting procedure is used to correct the raw $\pt$ spectrum.
Fig.~\ref{fig:efficiency} shows the $A \times \epsilon$  as a function of $\pt$ for minimum bias events in pp and in p--Pb collisions.
The drop in efficiency is seen at intermediate $\pt$ that arises due to the rejection of protons above $p > 1.1$ GeV/$c$ and kaons above \mbox{$p > 0.6$ GeV/$c$} when PID information is only available from the TPC. 
   Since no significant variation of $A \times \epsilon$ with event multiplicity was observed in p--Pb collisions, the $A \times \epsilon$ obtained in MB events was used for all multiplicity intervals to have a better statistical precision.
%
%
\subsection{Correction and normalization}
  
The final $\pt$ spectra were calculated from the raw yields as
\begin{equation}  
\frac{1}{N_{\rm evt}} \frac{{\rm d}^2N}{{\rm d}y{\rm d}\pt} = \frac{1}{N_{\rm evt}^{\rm trig}} \frac{N^{\rm raw}(\pt)}{\Delta y \ \Delta \pt}
\frac{\epsilon_{\rm trig}}{A \times \epsilon(\pt) \ BR} \ \frac{\epsilon_{\rm vert}}{\epsilon_{\rm sig}} \epsilon_{\rm G/F}(\pt) \ .
\end{equation}  
  
The raw yields ($N^{\rm raw}$) were corrected for $A \times \epsilon$ of the detectors, branching ratio (BR) of the decay channel, 
the trigger efficiency ($\epsilon_{\rm trig}$), GEANT3/FLUKA correction ($\epsilon_{\rm G/F}$), signal loss correction ($\epsilon_{\rm sig}$) and event loss correction due to the vertex reconstruction inefficiency ($\epsilon_{\rm vert}$). A GEANT3-based~\cite{Brun:1994aa} simulation of the ALICE detector response was used to correct the yields for both collision systems. 
The GEANT3 version used for correcting the yields from pp data overestimates the interactions of $\overline{\rm{p}}$ and K$^-$ with the material, especially at low $\pt$. Therefore the efficiency is scaled by a factor $\epsilon_{\rm G/F}$, estimated with a dedicated FLUKA simulation~\cite{Abelev:2013vea, Battistoni:2007zzb}.
In pp collisions, the yields are normalized to the number of inelastic collisions by applying a trigger efficiency correction ($\epsilon_{\rm trig}$) of  $0.85^{+0.06}_{-0.03}$ \cite{Adam:2015gka} to the total number of triggered events ($N_{\rm evt}^{\rm trig}$). The correction due to vertex reconstruction inefficiency ($\epsilon_{\rm vert}$) in pp collision is negligible ($\sim$0.1\%) and hence not applied.
For MB p--Pb collisions, the yields are normalized to the number of non-single diffractive (NSD) events after applying the correction factors of $\epsilon_{\rm trig}$ for event selection and $\epsilon_{\rm vert}$ due to the  primary vertex reconstruction inefficiency, resulting in a total scaling factor of 0.964 \cite{Adam:2016bpr}.
For the multiplicity dependent study of $\Ls$, the yields are normalized to the number of events in the respective V0A multiplicity event class.  Only events with a reconstructed primary vertex were considered in the computation of $A \times  \epsilon$.  Therefore, a correction for the vertex reconstruction inefficiency ($\epsilon_{\rm vert}$) has to be applied in each V0A multiplicity event class. The correction is about 0.95 for the 60--100\% multiplicity interval and is unity for other multiplicity intervals. The signal loss correction, $\epsilon_{\rm sig}$ is measured as a function of $\pt$ and it corresponds to the resonances that are not reconstructed in the events missing due to the trigger selection.  This correction is significant (of the order of few \%) at low $\pt$ ($<$ 2 GeV/$c$) in the lowest multiplicity interval for p--Pb collisions and in pp collisions. The values of $\epsilon_{\rm sig}$($\pt$) are negligible for other multiplicity intervals in p--Pb collisions.
 \subsection{Sources of systematic uncertainties}
The systematic uncertainties of the $\Ls$ yields are summarized in Table~\ref{table:systematic}. The main sources of systematic uncertainty are signal extraction, track selection cuts, PID selection cuts, global tracking efficiency, the material thickness traversed by the particles (material budget) and the hadronic interaction cross-section in the detector material. 
No event multiplicity dependence of systematic effects was observed in p--Pb data, thus the uncertainties estimated for minimum bias collisions were used for all multiplicity intervals. 
One of the main sources of systematic uncertainties is the raw yield extraction procedure. This contribution is labelled as ``Signal extraction'' and accounts for uncertainties mainly due to the choice of the background normalization region, the fitting range, the residual background shape and  
variation of the mass resolution in the fitting function. The normalization ranges have been varied between 1.7 GeV/$c^2$ and 2.2 GeV/$c^2$.
The lower mass limit of the fitting range was varied within  $\pm$20 MeV/$c^2$ while the higher mass limit was varied within $\pm$100 MeV/$c^2$ about the default fitting range of 1.45--1.65 GeV/$c^{2}$.
A second order polynomial is used as the default function to describe the residual background and a third order polynomial  is used to estimate the systematic uncertainty. The mass resolution was varied within the range of uncertainties observed in the simulation.
Track selection uncertainties were estimated to be 2\% for a single charged track in~\cite{ALICE:2012xs, Adam:2015qaa} and therefore correspond to a 4\% uncertainty for $\Ls$ decaying into two daughter particles.
In order to study the effect of the PID selection on signal extraction, the cuts on the TPC d$E$$\rm{/}$d$x$ and TOF time-of-flight values were varied by 1$\sigma$.
 This results in average uncertainties in the yields of 1.8\% and 2.1\% for pp and p--Pb collisions, respectively.  
The uncertainty on the determination of the global tracking efficiency (ITS-TPC matching uncertainty) is independent of $\pt$ and was evaluated to be 3\% for a single charged particle~\cite{ALICE:2012xs, Adam:2015qaa}, which results in a 6\% uncertainty when two tracks are combined in the invariant-mass analysis.
\begin{table*}[ht]
\caption{Sources of systematic uncertainties for $\Ls$ yields (d$^2N$/(d$y$d$\pt$)). For each source the average relative uncertainties are listed. }
\begin{tabular*}{\textwidth}{@{\extracolsep{\fill}}lrrrrrl@{}}
\hline
 \multicolumn{1}{c}{Source} & \multicolumn{2}{c}{p--Pb, $\sn$ $\mathrm{=}$ 5.02 TeV} & \multicolumn{2}{c}{pp, $\s$ $\mathrm{=}$ 7 TeV}  \\
\hline
\hline
Signal extraction       &   \multicolumn{2}{c}{ 5.0\% }      & \multicolumn{2}{c}{ 4.0\% }\\
Track selection cuts  &   \multicolumn{2}{c}{ 4.0\% }      & \multicolumn{2}{c}{ 4.0\% } \\
Particle identification &   \multicolumn{2}{c}{ 1.8\% }   & \multicolumn{2}{c}{ 2.1\% } \\
Global tracking efficiency & \multicolumn{2}{c}{ 6.0\% } & \multicolumn{2}{c}{ 6.0\% }  \\
Material budget         &   \multicolumn{2}{c}{ 1.5\% ($\pt < $ 3.5 GeV/$c$) }  & \multicolumn{2}{c}{ 1.4\% ($\pt <$ 3.5 GeV/$c$) } \\
Hadronic interaction  &   \multicolumn{2}{c}{ 3.3\%  ($\pt <$ 3.5 GeV/$c$) }  & \multicolumn{2}{c}{ 3.0\% ($\pt <$ 3.5 GeV/$c$) } \\
\hline
Total  &   \multicolumn{2}{c}{ 9.7\% }  & \multicolumn{2}{c}{ 9.1\% }\\ 
\hline
\hline
\end{tabular*}
\label{table:systematic}
\end{table*}
The systematic uncertainties in the $\Ls$ yield due to the material budget and hadronic interaction cross section in the detector material have been found to be constant up to $\pt$ $\mathrm{=}$ 3.5 GeV/$c$ and negligible at higher $\pt$. 
The uncertainties due to signal extraction and PID are uncorrelated with $\pt$, whereas the global tracking, track cuts, material budget and hadronic cross-section uncertainties are correlated with $\pt$. These 
$\pt$-correlated uncertainties cancel when calculating the uncertainties of the particle ratios.

\section{Results and discussion} \label{sec:ResultsandDiscussion} 
This section presents the results obtained for the $\Ls$, which include the $\pt$-differential spectra, integrated yields (d$N$$\rm{/}$d$y$) and mean transverse momentum ($\mpt$) values in MB pp collisions at $\s$ $\mathrm{=}$ 7 TeV and p--Pb collisions at $\sn$ $\mathrm{=}$  5.02 TeV in the NSD event class and different multiplicity intervals.
This section also describes a study of the radial flow effect in p--Pb collisions for the $\Ls$ $\pt$-spectra. 
The results are also used to study strangeness enhancement as a function of the charged-particle multiplicity at mid-pseudorapidity.
 \subsection{Transverse momentum spectra}
\begin{figure}[h] 
\begin{minipage}{\columnwidth}
\centering
\includegraphics[width=12cm]{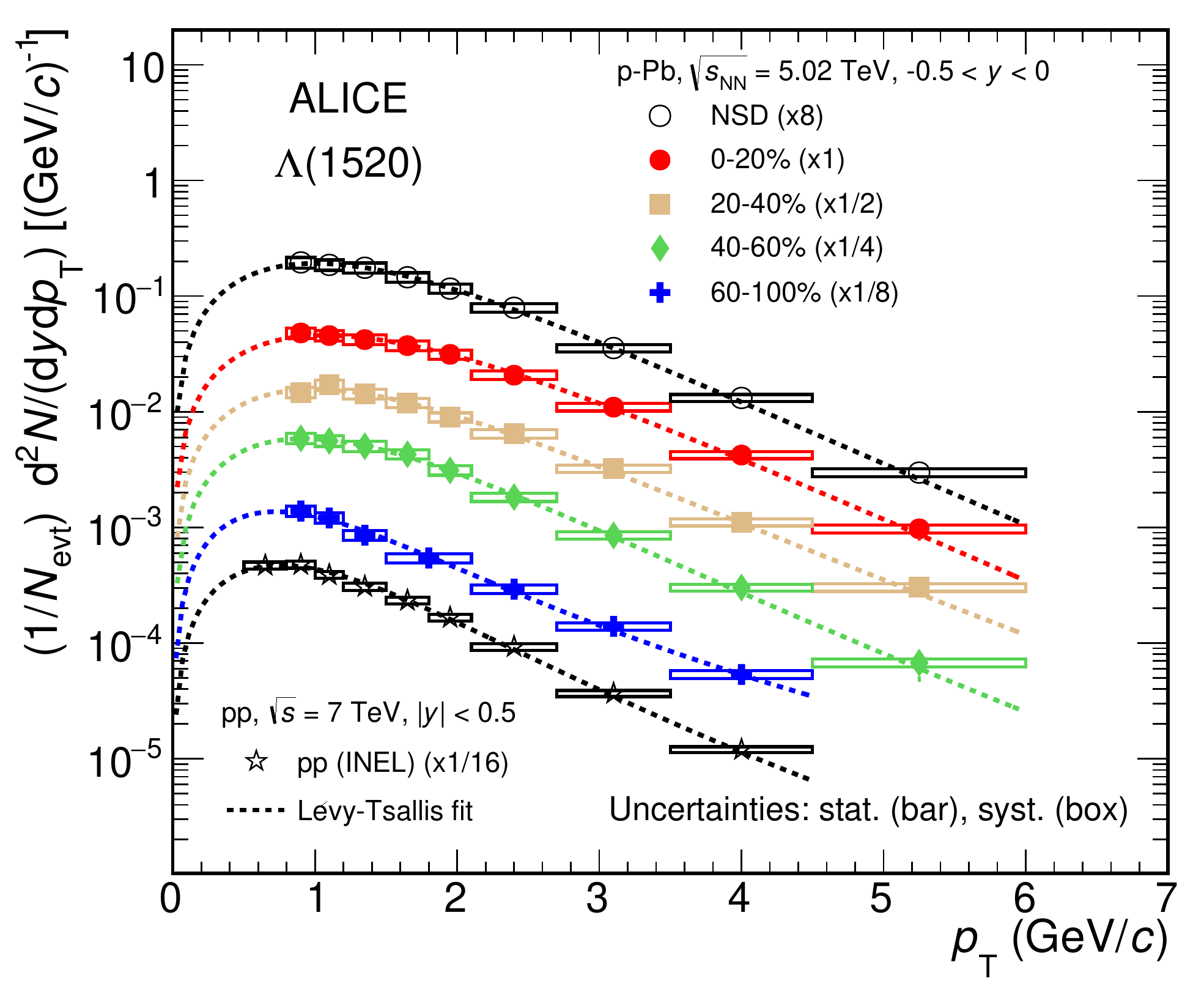} 
\end{minipage}
\caption{ (Color online) $\pt$ spectra of $\Ls$ measured with ALICE in the rapidity range $|y| < $ 0.5 in pp collisions at $\s$ $\mathrm{=}$ 7 TeV and in the rapidity range $-0.5 < y < 0$ in p--Pb collisions at \mbox{$\sn$ $\mathrm{=}$ 5.02 TeV} for minimum bias and different multiplicity intervals (V0A estimator). The multiplicity-dependent spectra are normalized to the visible cross-section, whereas the minimum bias spectrum is normalized to the fraction of NSD events. The minimum bias spectrum in pp collisions is normalized to the number of inelastic events. Statistical and systematic uncertainties ($\pt$-uncorrelated) are indicated as bars and boxes, respectively. Dashed lines represent L\'{e}vy-Tsallis fits.}
\label{fig:spectra}
\end{figure}

The $\pt$ spectra of $\Ls$ measured in the rapidity range $|y| < $ 0.5 in inelastic pp collisions and \mbox{$-0.5 < y <$  0} in p--Pb collisions for various event classes are shown in Fig. \ref{fig:spectra}. 
The L\'{e}vy-Tsallis parameterization~\cite{Tsallis:1987eu} is used to fit the $\pt$-differential spectra, d$^{2}N/({\rm d}p_{\rm T} {\rm d}y)$. 
The function provides a good description of the measured points over the whole $\pt$ range with a $\chi^2$/ndf less than 1.
The fits are used to extrapolate the spectra down to zero $\pt$ and to high $\pt$ (up to 10 GeV/$c$). 
The d$N$$\rm{/}$d$y$ and $\mpt$ are obtained from the spectra in the measured ranges and from the fits at lower and higher momenta.  The $\pt$-correlated uncertainties are not considered for fitting,  but propagated separately to the final results.
Table~\ref{table:yieldmpT} reports ${\rm d}N /{\rm d}y$ and $\mpt$ for $\Ls$ along with the extrapolation fraction and $\chi^2/$ndf values obtained from the fits.
The first uncertainty is statistical and the second one is the total systematic uncertainty. 
The extrapolation fraction in p--Pb collisions varies from 20\% in the 0--20\% multiplicity interval to 33\% in the 60--100\% multiplicity interval; in pp collisions it is below 16\%.
%
\begin{table*}[ht]
\caption{${\rm d}N /{\rm d}y$ and $\mpt$ along with the extrapolation fraction (Extr.) and $\chi^2/$ndf from the fit to the $\pt$ distribution. The values included in the table correspond to four multiplicity intervals (V0 estimator) and NSD events of p--Pb collisions at $\sn$ $\mathrm{=}$ 5.02 TeV and inelastic pp collisions at $\s$ $\mathrm{=}$ 7 TeV. The first uncertainty is the statistical and the second one is the total systematic uncertainty.}
\centering
\begin{tabular*}{\columnwidth}{@{\extracolsep{\fill}}cccccc@{}}
\hline
System & Event Class  &d$N/$d$y$ &$\mpt$ (GeV/$c$) & Extr. & $\chi^2/$ndf \\
\hline \hline
p--Pb &  0--20\% & 0.099 $\pm$ 0.004 $\pm$ 0.011 & 1.675 $\pm$ 0.036 $\pm$ 0.040 & 0.205 & 0.2 \\
	 & 20--40\%& 0.065 $\pm$ 0.002 $\pm$ 0.007 & 1.607 $\pm$ 0.032 $\pm$ 0.043 & 0.225 & 0.4 \\
	 & 40--60\%& 0.044 $\pm$ 0.002 $\pm$ 0.005 & 1.475 $\pm$ 0.034 $\pm$ 0.037 & 0.258 & 0.2 \\
	 &60--100\%&0.018 $\pm$ 0.001 $\pm$ 0.002 & 1.405 $\pm$ 0.093 $\pm$ 0.074 & 0.332 & 0.6 \\
\hline
p--Pb& NSD       &0.049 $\pm$ 0.001 $\pm$ 0.005 & 1.579 $\pm$ 0.020 $\pm$ 0.035 & 0.227 & 0.1 \\
\hline
pp	 & INEL       &0.012 $\pm$ 0.0003$\pm$ 0.0012& 1.273 $\pm$ 0.021 $\pm$ 0.043 & 0.156 & 0.4 \\
\hline \hline
\end{tabular*}
\label{table:yieldmpT}
\end{table*}
The systematic uncertainties on ${\rm d}N /{\rm d}y$ are dominated by the $\pt$-uncorrelated uncertainties of the measured spectra 
(about 7.6\% in p--Pb and 6.4\% in pp), the $\pt$-correlated contributions from the global tracking efficiency (6\% both in pp and p--Pb)~\cite{Adam:2015qaa, ALICE:2012xs} and the extrapolation of the yield (2.1--2.9\% in p--Pb and  2.0\% in pp). The uncertainties due to the extrapolation of the yields are calculated using different functions: Blast-Wave~\cite{Schnedermann:1993ws}, $m_{\rm T}$-exponential, Boltzmann and Fermi-Dirac functions~\cite{Abelev:2008ab} and the standard deviation of the ${\rm d}N /{\rm d}y$ values calculated using
these functions is used as the systematic uncertainty.  
Similarly, the major contributors to the systematic uncertainty of $\mpt$ are the $\pt$-uncorrelated systematic uncertainties on the measured $\pt$-differential yields (about 2.6\% in p--Pb and 2.3\% in pp) and the standard deviation of the $\mpt$ values calculated using the different extrapolation functions. The $\pt$-correlated uncertainties are not included in the $\mpt$ as the correlated uncertainties act as the normalization constant and do not affect the spectral shape. 
\subsection{ Average transverse momentum and mass ordering}
\begin{figure}[!ht] 
\begin{minipage}{\columnwidth}
\centering
\includegraphics[width=0.5\textwidth]{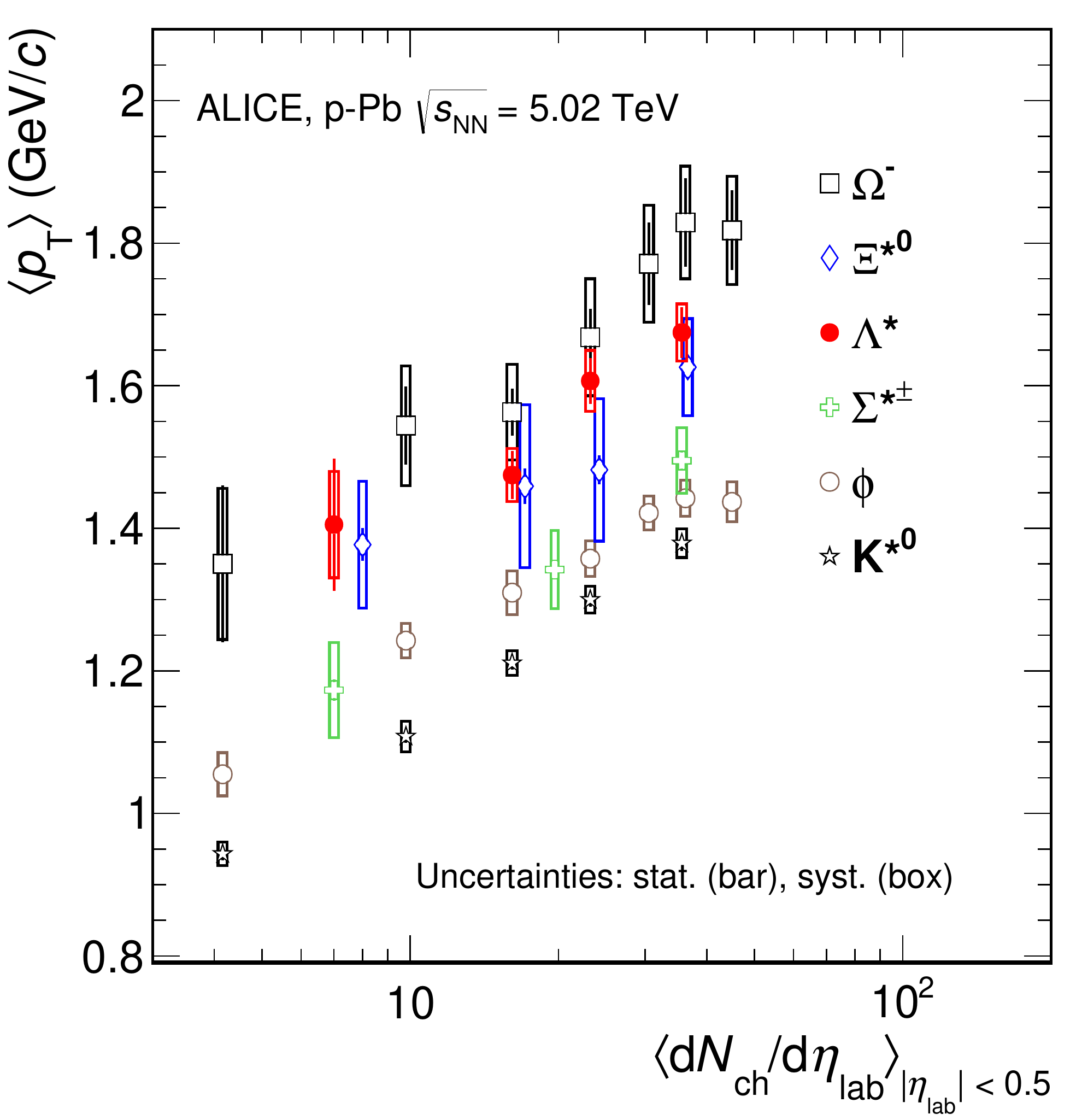} 
\end{minipage}
\caption{ (Color online) The $\mpt$ of $\Ls$ compared with previously measured $\mpt$ values of K$^{*0}$, ${\rm \phi}$, ${\rm \Sigma^{*\pm}}$, ${\rm\Xi^{*0}}$  and ${\rm\Omega^-}$ in p--Pb collisions at $\sn$ $\mathrm{=}$ 5.02 TeV as a function of the mean charged-particle multiplicity density $\dnchdeta$, measured in the pseudorapidity range $|\eta_{\rm lab}| < 0.5$~\cite{Adam:2016bpr, Adam:2015vsf, Adamova:2017elh}. The ${\rm\Xi^{*0}}$ points are slightly displaced along the abscissa for clarity. Statistical uncertainties are represented as bars, whereas boxes indicate systematic uncertainties.}
\label{fig:meanptdndch}
%
\begin{minipage}{\columnwidth}
\centering
\includegraphics[width=0.5\textwidth]{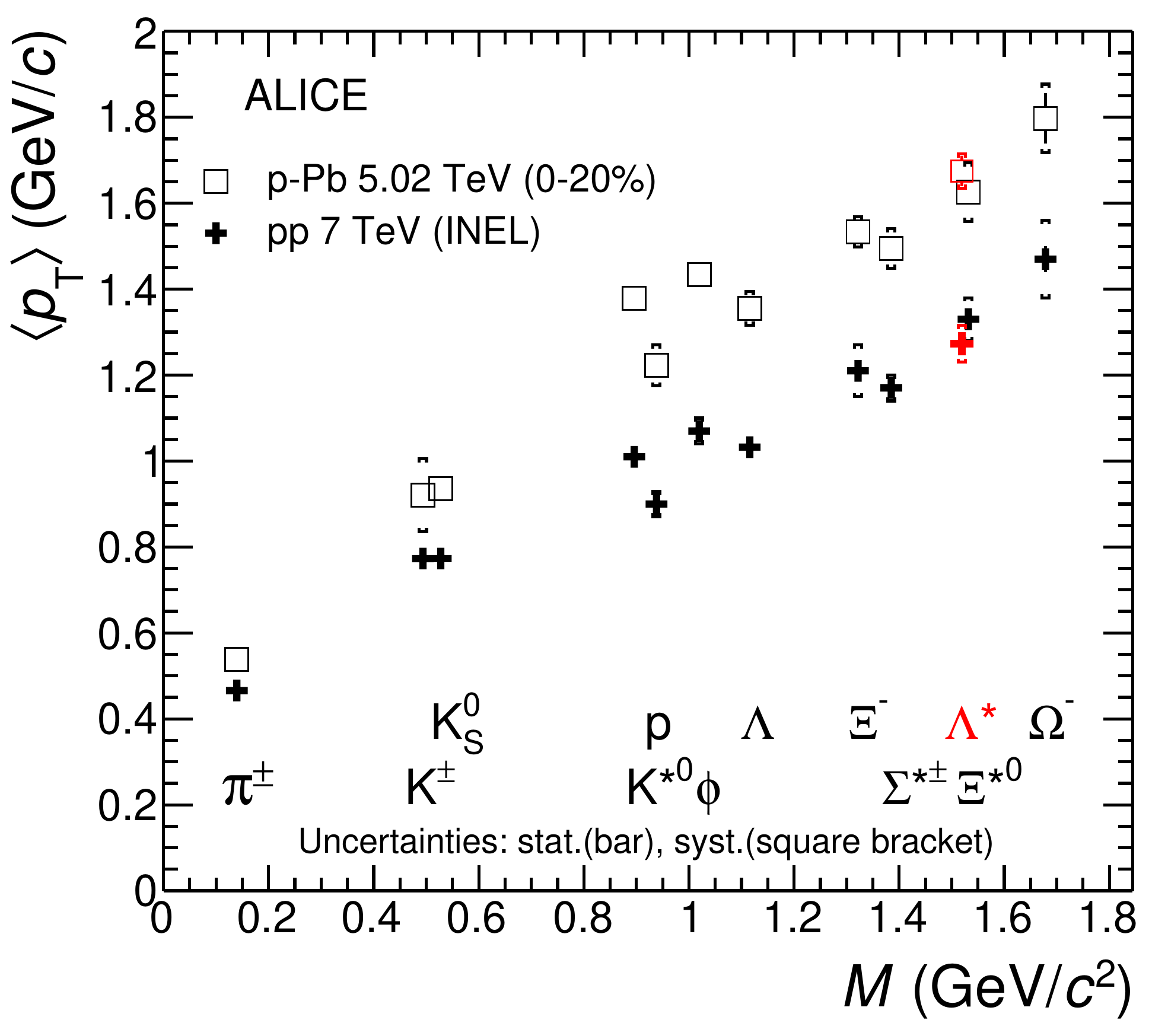} 
\end{minipage}
\caption{ Mass dependence of the $\mpt$ of identified particles measured  in inelastic pp collisions at \mbox{$\s$ $\mathrm{=}$ 7 TeV}~\cite{Adam:2015qaa, Abelev:2012hy, Abelev:2014qqa} and in p--Pb collisions at $\sn$ $\mathrm{=}$ 5.02 TeV for the 0--20\% multiplicity intervals~\cite{Abelev:2013haa, Adam:2015vsf, Adamova:2017elh}. Statistical uncertainties are represented as bars, square brackets indicate total systematic uncertainties.}
\label{fig:meanptmass}
\end{figure}

In high-energy heavy-ion collisions, the expansion velocity of the medium drives the spectral shapes of final-state particles.
If an increase in $\mpt$ with the mean charged-particle multiplicity density is observed for different particles, then it may suggest collective (hydrodynamic) behavior of the system~\cite{Abelev:2013haa, Abelev:2013vea}.
Figure~\ref{fig:meanptdndch} shows the $\mpt$ of different particle species as a function of the mean
charged-particle multiplicity density $\dnchdeta$ within $|\eta_{\rm{lab}}| < 0.5$. 
%
The values for the $\Ls$ are compared with those of other hyperons and mesons observed in p--Pb collisions at $\sn$ $\mathrm{=}$ 5.02 TeV; such as K$^{*0}$, ${\rm \phi}$, ${\rm \Sigma^{*\pm}}$, ${\rm\Xi^{*0}}$, ${\rm\Omega^-}$~\cite{ Adam:2016bpr, Adam:2015vsf, Adamova:2017elh}.
An increasing trend of $\mpt$ from low to high multiplicity is observed for all particles. 
This enhancement in $\mpt$ of the $\Ls$ is consistent, within uncertainties, with that observed for other hadrons as shown in the figure. The change of the $\mpt$ for the K$^{*0}$ meson is the largest, which may be due to the suppression of low $\pt$ particles due to re-scattering in the hadronic medium for the higher multiplicity intervals.
A mass ordering of $\mpt$ is observed among light-flavor baryons, including the $\Ls$. 
This mass hierarchy including other light flavored hadrons is shown in Fig.~\ref{fig:meanptmass}.
The figure shows the $\mpt$ of several hadron species as a function of mass for inelastic pp collisions at $\s$ $\mathrm{=}$ 7 TeV~\cite{Adam:2015qaa, Abelev:2012hy, Abelev:2014qqa} and for the 0--20\% multiplicity interval of p--Pb collisions at $\sn$ $\mathrm{=}$ 5.02 TeV~\cite{Abelev:2013haa, Adam:2015vsf, Adamova:2017elh}. The $\mpt$ increases with increasing mass in both collision systems. The baryonic resonance $\Ls$ follows the same trend as the other baryons. We observe two different trends in the $\mpt$  for mesons and baryons. These results reflect the violation of mass ordering in the $\mpt$  of the produced particles in pp and p--Pb collisions.

%
%
%
 \subsection{Collective radial expansion}
In heavy-ion collisions, the flattening of transverse momentum distributions of hadrons and their $\mpt$ ordering with mass is explained by the collective radial expansion of the system~\cite{Heinz:2004qz}. 
\begin{figure}[!ht] 
\begin{minipage}{\columnwidth}
\centering
\includegraphics[width=8cm]{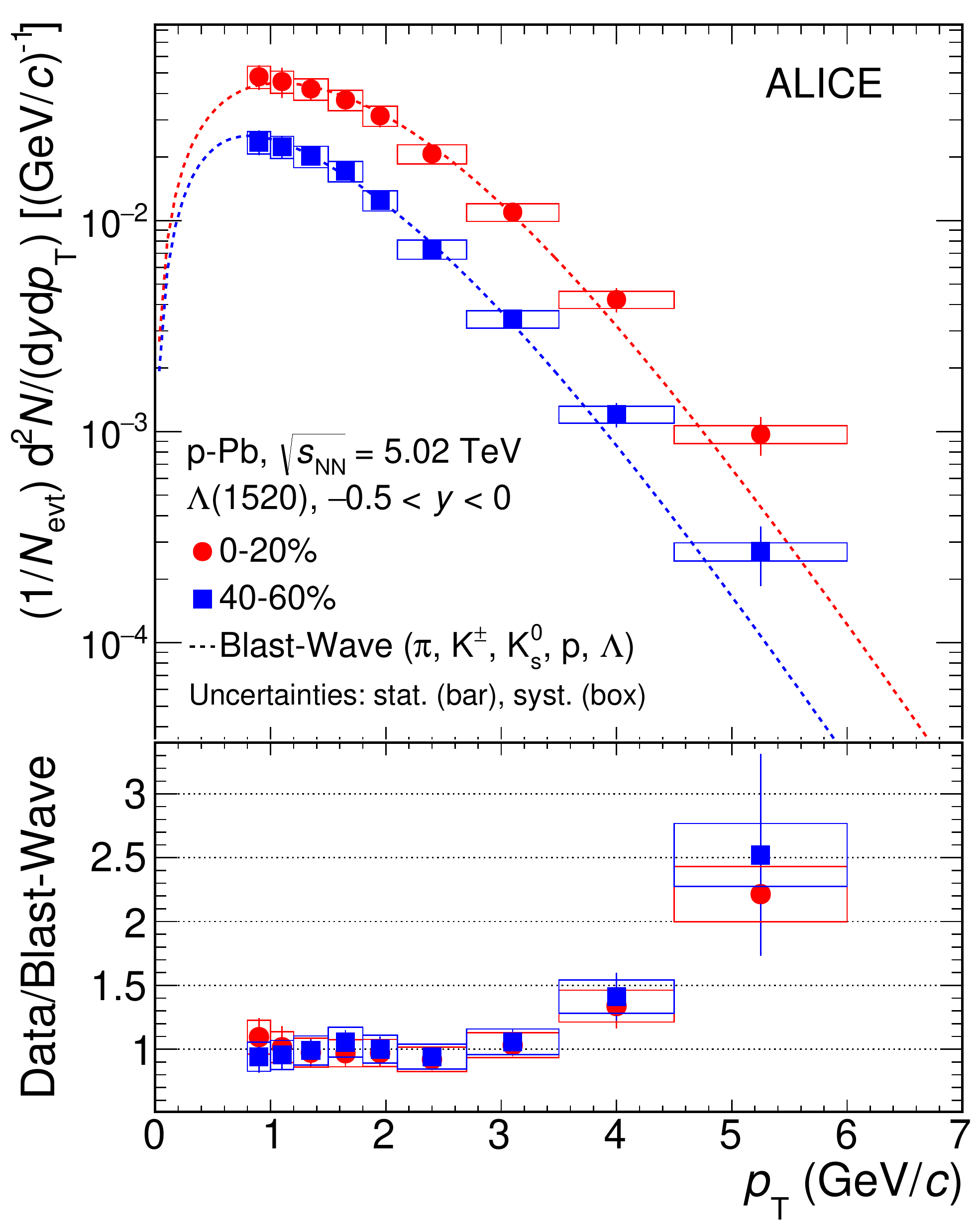} 
\end{minipage}
\caption{ (Color online) The upper panel shows the $\pt$ distribution of $\Ls$ in p--Pb collisions at \mbox{$\sn$ $\mathrm{=}$ 5.02 TeV} in the 0--20\% (red) and 40--60\% (blue) multiplicity intervals. The dashed curves represent predictions from the Blast-Wave model~\cite{Abelev:2013haa}, where the parameters are obtained from simultaneous fits to $\pi^{\pm}$, K$^{\pm}$, K$_{\rm{S}}^{0}$, p($\overline{\rm p}$) and $\Lam$($\overline{\Lam}$)  $\pt$ spectra and the shapes are normalized to the data. The lower panel shows the ratios of the measured distributions to the values from the respective Blast-Wave functions. The statistical uncertainties are shown as bars and the systematic uncertainties are shown as boxes. }
\label{fig:BWradialflow}
\end{figure}
%
%
Previously, the Blast-Wave formalism~\cite{Schnedermann:1993ws} successfully described simultaneously the $\pi^{\pm}$, K$^{\pm}$, K$_{\rm{S}}^{0}$, p($\overline{\rm p}$) and $\Lam$($\overline{\Lam}$) spectra in different multiplicity intervals of p--Pb collisions~\cite{Abelev:2013haa}. 
The $\pt$ ranges for the simultaneous fit of $\pi^{\pm}$, K$^{\pm}$, K$_{\rm{S}}^{0}$, p($\overline{\rm p}$) and $\Lam$($\overline{\Lam}$) are 0.5$-$1 GeV/$c$, 0.2$-$1.5 GeV/$c$, 0$-$1.5 GeV/$c$, 0.3$-$3 GeV/$c$ and \mbox{0.6$-$3 GeV/$c$}, respectively.
The Blast-Wave model is not expected to be valid at high $\pt$.
Here we have used the same Blast-Wave parameters as extracted in Ref.~\cite{Abelev:2013haa} to obtain a prediction for the $\pt$ spectra of the $\Ls$. The parameters are listed in Table~\ref{table:bwParam}.
The predicted Blast-Wave shapes are normalized to the data in a momentum range up to 3 GeV/$c$ in the 0--20\% and 40--60\% multiplicity intervals.
The results are shown in the upper panel of Fig.~\ref{fig:BWradialflow}.  The lower panel of the same figure shows the ratio of the measured data points and the corresponding values from the functions.
This shows that the Blast-Wave function describes the shape of the $\Ls$ spectra well up to $\pt$ $\mathrm{=}$ 3.5 GeV/$c$. Although the Blast-Wave function underpredicts the $\pt$-differential yields at high $\pt$, the ratio of the measurement to the function is found to be independent of event multiplicity in the measured region. 
This suggests that (1) the $\Ls$ $\pt$ spectra agree with the Blast-Wave shape constrained by other light hadrons (at least in the $\pt$ range where the Blast-Wave model is expected to work); and (2) the $\Ls$ participates in the same collective radial flow as the other light hadrons. The measurements for the $\Ls$ are also consistent with an increase of radial flow with multiplicity. It may be noted that the $\Ls$ spectral shape and $\mpt$ were also found to be consistent with a hydrodynamic evolution in Pb--Pb collisions at $\sn$ $\mathrm{=}$  2.76 TeV~\cite{ALICE:2018ewo}.

\setlength{\tabcolsep}{28pt}
\begin{table*}
\caption{Blast wave parameters from fits to $\pi^{\pm}$, K$^{\pm}$, K$_{\rm{S}}^{0}$, p($\overline{\rm p}$) and $\Lam$($\overline{\Lam}$)  $\pt$ spectra in p--Pb collisions at $\sn$ $\mathrm{=}$  5.02 TeV~\cite{Abelev:2013haa}.}
\begin{tabular}{ c c c c }
\hline\hline 
Centrality & $T_{\mathrm{kin}}$ (GeV) & $\beta_{\mathrm{s}}$ & $n$ \\\hline 
0--20\% & $0.147 \pm 0.005$  & $0.833 \pm 0.0095 $ & $1.16 \pm 0.035$  \\\hline 
40--60\% & $0.164 \pm 0.004$  & $0.435 \pm 0.011 $ & $1.73 \pm 0.07$   \\\hline \hline 
\end{tabular}
\label{table:bwParam}
\end{table*}

\subsection{Ratios of integrated yields }
The ratios of the yields of particles with varying strangeness content, mass and lifetime are key observables in the study of particle production mechanisms. 
Short-lived particles such as $\Ls$, ${\rm \Xi^{*0}}$, K$^{*0}$ and $\phi$ are used to extract information on the lifetime of the hadronic phase in heavy-ion collisions and on mechanisms, such as re-scattering and regeneration, which affect resonance yields before kinetic freeze-out (when the constituents of the system cease to interact elastically).
%
\begin{figure}[h] 
\begin{minipage}{\columnwidth}
\centering
\includegraphics[width=7.5cm]{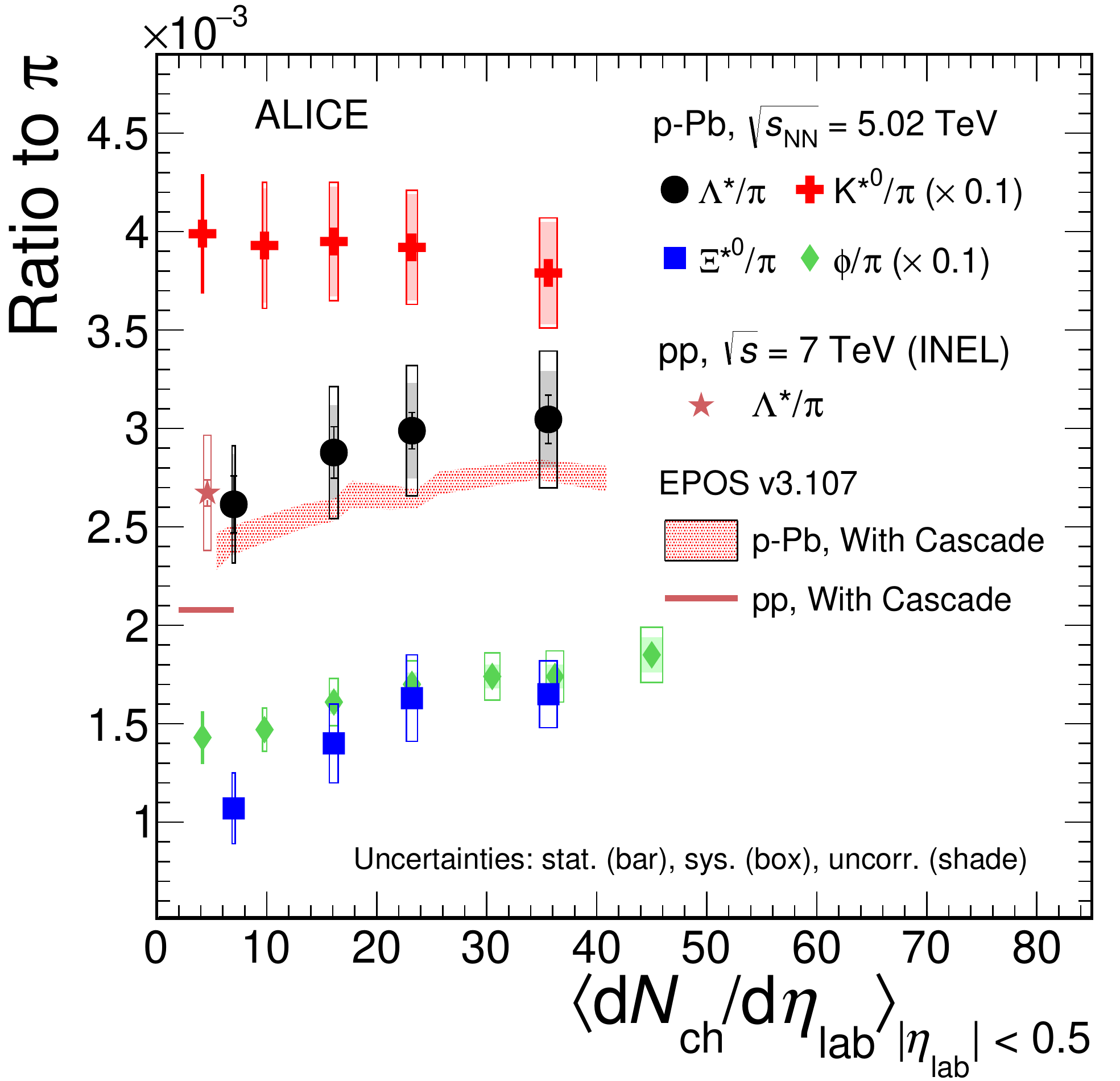} 
\includegraphics[width=7.5cm]{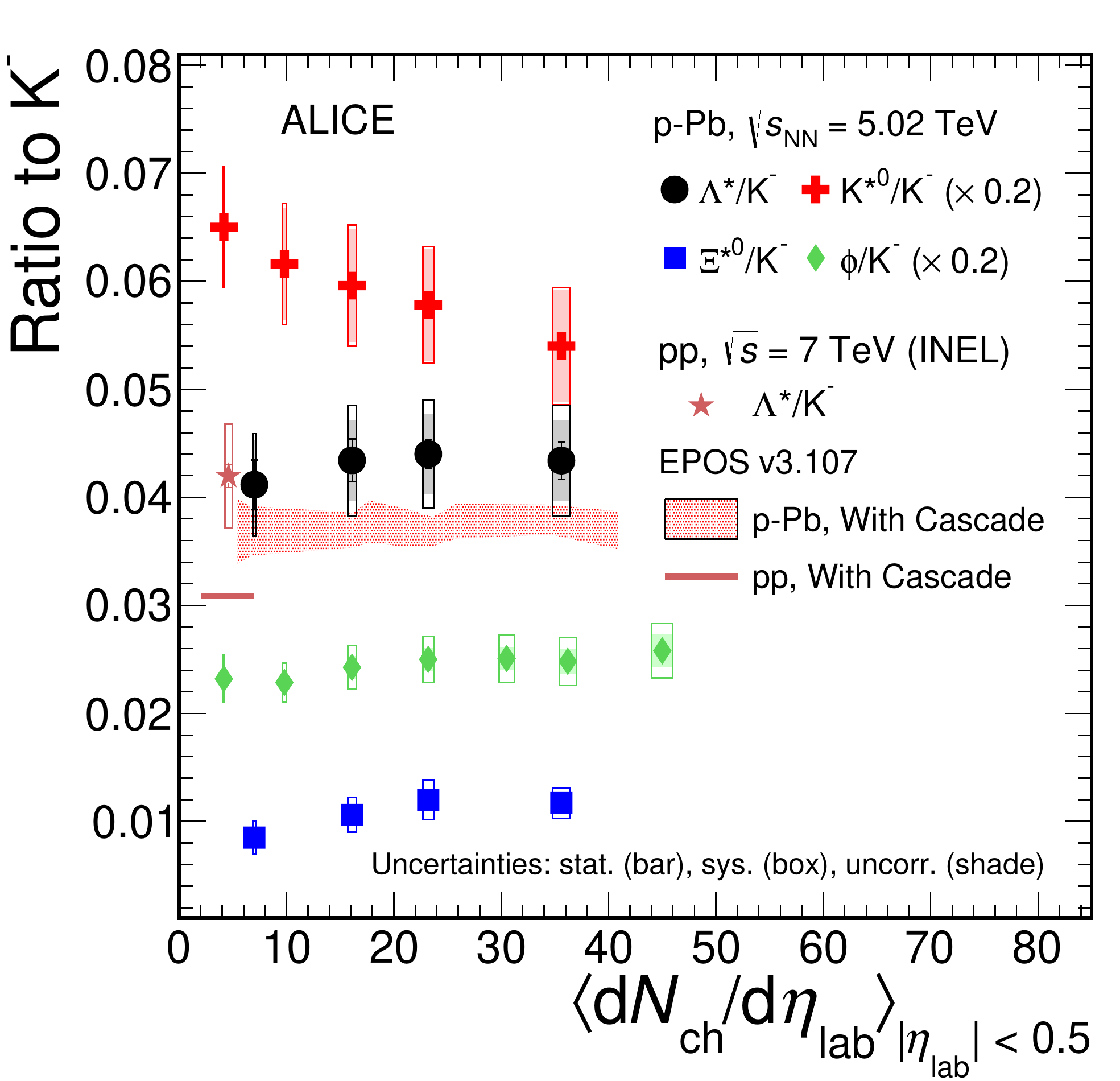} 
\end{minipage}
\caption{ (Color online) Ratio of $\Ls$, K$^{*0}$, ${\rm \Xi^{*0}}$ and ${\rm \phi}$ to charged $\pi$ (left) and K$^-$ (right) in p--Pb collisions at $\sn$ $\mathrm{=}$ 5.02 TeV as a function of the average charged-particle density $\dnchdeta$ measured at midrapidity~\cite{Adam:2016bpr, Adamova:2017elh}. Statistical uncertainties (bars) are shown together with total (hollow boxes) and multiplicity-uncorrelated (shaded boxes) systematic uncertainties. The EPOS3 model predictions~\cite{Drescher:2000ha, Werner:2010aa, Werner:2013tya, Bass:1998ca, Knospe:2015nva} with statistical uncertainties are shown as horizontal bars (pp collisions at $\s$ $\mathrm{=}$ 7 TeV) and shaded bands (p--Pb collisions at $\sn$ $\mathrm{=}$ 5.02 TeV).}
\label{fig:Ratiotopion}
\end{figure}
%
%
In Fig.~\ref{fig:Ratiotopion}, the yield ratios of resonances to charged $\pi$ and K$^{-}$ are compared. The yield ratios of strange resonances to the non-strange $\pi$ show a rise from low to high multiplicities (except for K$^{*0}/\pi$).
The $\Ls/\pi$ ratio shows a hint of an increase with increasing charged-particle multiplicity, but due to the large uncertainties a strong conclusion cannot be drawn. This enhancement is more prominent for the ${\rm \Xi^{*0}}/\pi$ (the ratio of a doubly strange particle to a non-strange particle) and $\phi/\pi$ ratios. 
This yield enhancement is typically attributed to a reduced canonical suppression of strangeness production in larger freeze-out volumes~\cite{Sollfrank:1997bq} or to an enhanced strangeness production in a quark--gluon plasma~\cite{Letessier:2006wn}. In contrast, the K$^{*0}/\pi$ ratio shows no enhancement. This is connected to the negative slope of the K$^{*0}/$K$^-$ ratio (see the right panel of Fig.~\ref{fig:Ratiotopion}) which may hint at the suppression of K$^{*0}$ yields due to re-scattering effects in the hadronic medium in p--Pb collisions~\cite{Adam:2016bpr}.  The $\Ls/$K$^-$ ratio shows no change with increasing charged-particle multiplicity and the ratio is consistent with the value measured in pp collisions. Similar behaviour is seen for the $\phi/$K$^-$ ratio. 
%
The measured ratios are compared with the results from the EPOS3 model (version v3.107) with a hadronic cascade phase (EPOS3 + UrQMD). The EPOS3 event generator is based on a 3+1D viscous hydrodynamical evolution~\cite{Drescher:2000ha, Werner:2010aa, Werner:2013tya}. The initial conditions are described by the Gribov-Regge multiple scattering framework. The reaction volume consists of two parts: ``core'' and ``corona''. The core part constitutes the bulk matter simulated using 3+1D viscous hydrodynamics which thermalizes, flows and hadronizes. The corona part constitutes the hadrons from the string decays. These hadrons from core and corona part are fed into UrQMD~\cite{Bass:1998ca}, which includes the rescattering and regeneration effects. For pp collisions, EPOS3 under-predicts the data (horizontal bars), however for p--Pb collisions, EPOS3 agrees with the data (red shaded band) within uncertainties. EPOS3 also predicts a rise in the $\Ls/\pi$ ratio with increasing charged-particle multiplicity.

\begin{figure}[h!] 
\begin{minipage}{\columnwidth}
\centering
\includegraphics[width=8.0cm]{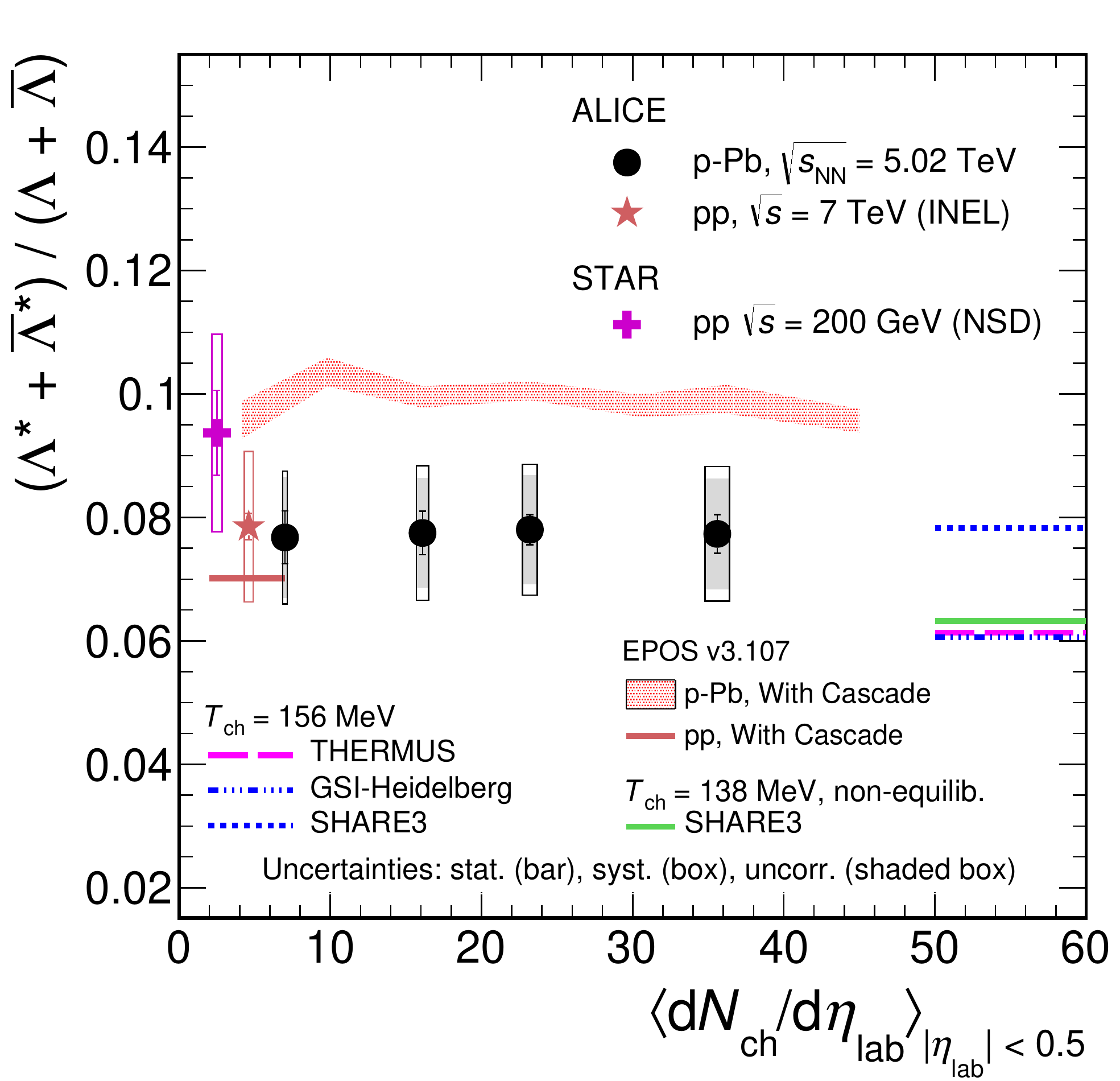} 
\end{minipage}
\caption{ (Color online) Ratio of $\Ls$ to $\Lam$ in inelastic pp collisions at $\s$ $\mathrm{=}$ 7 TeV, p--Pb collisions at \mbox{$\sn$ $\mathrm{=}$ 5.02 TeV}~\cite{Abelev:2013haa} and in NSD pp collisions at $\s$ $\mathrm{=}$ 200 GeV~\cite{Adams:2006yu, Abelev:2006cs}. Statistical uncertainties (bars) are shown together with total (hollow boxes) and multiplicity-uncorrelated (shaded boxes) systematic uncertainties. The results are compared with several model calculations (see text for details).} 
\label{fig:Ratiotolambda}
\end{figure}
Figure~\ref{fig:Ratiotolambda} shows the yield ratio of the $\Ls$ to the ground state $\Lam$ as a function of the average charged-particle density, $\dnchdeta$, measured at midrapidity. 
The figure shows the ratio for inelastic pp collisions at $\s$ $\mathrm{=}$ 7 TeV and $\s$ $\mathrm{=}$ 200 GeV (STAR Collaboration~\cite{Adams:2006yu, Abelev:2006cs}) and for p--Pb collisions at \mbox{$\sn$ $\mathrm{=}$ 5.02 TeV}~\cite{Abelev:2013haa} as a function of charged particle multiplicity.  The STAR measurement is consistent with the ALICE results within uncertainties. The $\Ls/\Lam$ ratio does not change with increasing charged-particle multiplicity.  This is in contrast to the measurements in Pb--Pb collisions at \mbox{$\sn$ $\mathrm{=}$ 2.76 TeV}~\cite{ALICE:2018ewo} where the $\Ls/\Lam$ ratio is observed to be suppressed in central (0--20\%) relative to peripheral (50--80\%) Pb--Pb collisions. 
The constant behavior of the $\Ls/\Lam$ ratios presented here along with those for ${\rm \Sigma^{*\pm}/\Lambda}$ and ${\rm \Xi^{*0}/\Xi^{-}}$ ratios reported in~\cite{Adamova:2017elh} as a function of $\dnchdeta$ indicates that the strangeness enhancement observed in p--Pb collisions depends predominantly on the strangeness content, rather than on the hyperon mass.
The $\Ls/\Lam$ ratios in pp and p--Pb collisions reported here are about a factor 2 higher than those measured in central Pb--Pb collisions.
THERMUS~\cite{Wheaton:2004vg}, GSI-Heidelberg~\cite{Andronic:2005yp} and SHARE3~\cite{Petran:2013dva} model calculations are able to describe the value of the ratio within 1.5 times the experimental uncertainty. 
The predictions of the three thermal models are for Pb--Pb collisions at the chemical freeze-out temperature (when the constituents of the system cease to interact inelastically) \mbox{$T_{\rm ch}$ $\mathrm{=}$ 156 MeV} and zero baryochemical potential. Among these models, SHARE3 (equilibrium) at \mbox{$T_{\rm ch}$ $\mathrm{=}$ 156 MeV} provides the best agreement with the data. EPOS3 with UrQMD agrees with the data qualitatively and indicates that the $\Ls$/$\Lam$ ratio is independent of the average charged-particle density in p--Pb collisions. 
This is in contrary to the observation that in central Pb--Pb collisions the $\Ls/\Lam$ ratio is suppressed compared to pp, p--Pb, peripheral (50--80\%) Pb--Pb collisions and thermal model calculations. The suppression of the $\Ls/\Lam$ ratio is consistent with the formation of a dense hadronic phase and re-scattering effects in central Pb--Pb collisions.
\section{Summary and conclusions} \label{sec:SummaryandConclusion}
The transverse momentum spectra of the $\Ls$ in pp collisions at $\s$ $\mathrm{=}$ 7 TeV and in p--Pb collisions at \mbox{$\sn$ $\mathrm{=}$  5.02 TeV}  have been measured using the ALICE detector in the rapidity ranges $|y| < 0.5$ and $-0.5 < y < 0$, respectively. 
The $\mpt$ of this baryonic resonance increases with the mean charged-particle multiplicity measured at mid-pseudorapidity ($|\eta_{\rm{lab}}| < 0.5$) and exhibits mass ordering when compared with other baryons (${\rm \Sigma^{*\pm}}$, ${\rm\Xi^{*0}}$, ${\rm\Omega^-}$ etc.).
The $\Ls/\pi$ ratio may exhibit an enhancement consistent with that observed for other strange hadrons~\cite{ALICE:2017jyt}.
The $\Ls$ $\pt$ spectra agree with the Blast-Wave shape constrained using other light hadrons, which, in the context of this particular model, can be interpreted as the $\Ls$ participating in the same collective radial flow as the other light hadrons.
The ratio of the $\Ls$ to the ground-state $\Lam$ shows no change with increasing charged-particle multiplicity in p--Pb collisions. This ratio is consistent with the measurement in pp collisions at $\s$ $\mathrm{=}$ 7 TeV and \mbox{200 GeV} within uncertainties. 
This measurement may indicate that the cumulative effect from the hadronic phase in p--Pb collisions is not enough to have significant influence on the $\Ls$ yield. 

The current measurements represent a useful baseline for the results in Pb--Pb collisions~\cite{ALICE:2018ewo}. The measurements of the $\Ls/\Lam$ ratio in pp, 
p--Pb, and peripheral Pb--Pb collisions indicate that the re-scattering effect plays an important role in central Pb--Pb collisions at $\sn$ $\mathrm{=}$ 2.76 TeV. A complete set of such measurements for many resonances with different lifetimes will allow the properties of the hadronic phase to be studied in more detail.

\newenvironment{acknowledgement}{\relax}{\relax}
    \begin{acknowledgement}
      \section*{Acknowledgements}

The ALICE Collaboration would like to thank all its engineers and technicians for their invaluable contributions to the construction of the experiment and the CERN accelerator teams for the outstanding performance of the LHC complex.
The ALICE Collaboration gratefully acknowledges the resources and support provided by all Grid centres and the Worldwide LHC Computing Grid (WLCG) collaboration.
The ALICE Collaboration acknowledges the following funding agencies for their support in building and running the ALICE detector:
A. I. Alikhanyan National Science Laboratory (Yerevan Physics Institute) Foundation (ANSL), State Committee of Science and World Federation of Scientists (WFS), Armenia;
Austrian Academy of Sciences, Austrian Science Fund (FWF): [M 2467-N36] and Nationalstiftung f\"{u}r Forschung, Technologie und Entwicklung, Austria;
Ministry of Communications and High Technologies, National Nuclear Research Center, Azerbaijan;
Conselho Nacional de Desenvolvimento Cient\'{\i}fico e Tecnol\'{o}gico (CNPq), Financiadora de Estudos e Projetos (Finep), Funda\c{c}\~{a}o de Amparo \`{a} Pesquisa do Estado de S\~{a}o Paulo (FAPESP) and Universidade Federal do Rio Grande do Sul (UFRGS), Brazil;
Ministry of Education of China (MOEC) , Ministry of Science \& Technology of China (MSTC) and National Natural Science Foundation of China (NSFC), China;
Ministry of Science and Education and Croatian Science Foundation, Croatia;
Centro de Aplicaciones Tecnol\'{o}gicas y Desarrollo Nuclear (CEADEN), Cubaenerg\'{\i}a, Cuba;
Ministry of Education, Youth and Sports of the Czech Republic, Czech Republic;
The Danish Council for Independent Research | Natural Sciences, the Carlsberg Foundation and Danish National Research Foundation (DNRF), Denmark;
Helsinki Institute of Physics (HIP), Finland;
Commissariat \`{a} l'Energie Atomique (CEA), Institut National de Physique Nucl\'{e}aire et de Physique des Particules (IN2P3) and Centre National de la Recherche Scientifique (CNRS) and R\'{e}gion des  Pays de la Loire, France;
Bundesministerium f\"{u}r Bildung und Forschung (BMBF) and GSI Helmholtzzentrum f\"{u}r Schwerionenforschung GmbH, Germany;
General Secretariat for Research and Technology, Ministry of Education, Research and Religions, Greece;
National Research, Development and Innovation Office, Hungary;
Department of Atomic Energy Government of India (DAE), Department of Science and Technology, Government of India (DST), University Grants Commission, Government of India (UGC) and Council of Scientific and Industrial Research (CSIR), India;
Indonesian Institute of Science, Indonesia;
Centro Fermi - Museo Storico della Fisica e Centro Studi e Ricerche Enrico Fermi and Istituto Nazionale di Fisica Nucleare (INFN), Italy;
Institute for Innovative Science and Technology , Nagasaki Institute of Applied Science (IIST), Japanese Ministry of Education, Culture, Sports, Science and Technology (MEXT) and Japan Society for the Promotion of Science (JSPS) KAKENHI, Japan;
Consejo Nacional de Ciencia (CONACYT) y Tecnolog\'{i}a, through Fondo de Cooperaci\'{o}n Internacional en Ciencia y Tecnolog\'{i}a (FONCICYT) and Direcci\'{o}n General de Asuntos del Personal Academico (DGAPA), Mexico;
Nederlandse Organisatie voor Wetenschappelijk Onderzoek (NWO), Netherlands;
The Research Council of Norway, Norway;
Commission on Science and Technology for Sustainable Development in the South (COMSATS), Pakistan;
Pontificia Universidad Cat\'{o}lica del Per\'{u}, Peru;
Ministry of Science and Higher Education and National Science Centre, Poland;
Korea Institute of Science and Technology Information and National Research Foundation of Korea (NRF), Republic of Korea;
Ministry of Education and Scientific Research, Institute of Atomic Physics and Ministry of Research and Innovation and Institute of Atomic Physics, Romania;
Joint Institute for Nuclear Research (JINR), Ministry of Education and Science of the Russian Federation, National Research Centre Kurchatov Institute, Russian Science Foundation and Russian Foundation for Basic Research, Russia;
Ministry of Education, Science, Research and Sport of the Slovak Republic, Slovakia;
National Research Foundation of South Africa, South Africa;
Swedish Research Council (VR) and Knut \& Alice Wallenberg Foundation (KAW), Sweden;
European Organization for Nuclear Research, Switzerland;
Suranaree University of Technology (SUT), National Science and Technology Development Agency (NSDTA) and Office of the Higher Education Commission under NRU project of Thailand, Thailand;
Turkish Atomic Energy Agency (TAEK), Turkey;
National Academy of  Sciences of Ukraine, Ukraine;
Science and Technology Facilities Council (STFC), United Kingdom;
National Science Foundation of the United States of America (NSF) and United States Department of Energy, Office of Nuclear Physics (DOE NP), United States of America.   
      
    \end{acknowledgement}
    \bibliographystyle{utphys}
    \bibliography{LambdaStarPaperDraft}

\providecommand{\href}[2]{#2}\begingroup\raggedright\begin{thebibliography}{10}

\bibitem{Adams:2005dq}
{\bfseries STAR} Collaboration, J.~Adams {\em et~al.}, ``{Experimental and
  theoretical challenges in the search for the quark gluon plasma: The STAR
  Collaboration's critical assessment of the evidence from RHIC collisions}'',
  \href{http://dx.doi.org/10.1016/j.nuclphysa.2005.03.085}{{\em Nucl. Phys.}
  {\bfseries A757} (2005) 102--183},
\href{http://arxiv.org/abs/nucl-ex/0501009}{{\ttfamily arXiv:nucl-ex/0501009
  [nucl-ex]}}.

\bibitem{Adcox:2004mh}
{\bfseries PHENIX} Collaboration, K.~Adcox {\em et~al.}, ``{Formation of dense
  partonic matter in relativistic nucleus-nucleus collisions at RHIC:
  Experimental evaluation by the PHENIX collaboration}'',
  \href{http://dx.doi.org/10.1016/j.nuclphysa.2005.03.086}{{\em Nucl. Phys.}
  {\bfseries A757} (2005) 184--283},
\href{http://arxiv.org/abs/nucl-ex/0410003}{{\ttfamily arXiv:nucl-ex/0410003
  [nucl-ex]}}.

\bibitem{Arsene:2004fa}
{\bfseries BRAHMS} Collaboration, I.~Arsene {\em et~al.}, ``{Quark gluon plasma
  and color glass condensate at RHIC? The Perspective from the BRAHMS
  experiment}'', \href{http://dx.doi.org/10.1016/j.nuclphysa.2005.02.130}{{\em
  Nucl. Phys.} {\bfseries A757} (2005) 1--27},
\href{http://arxiv.org/abs/nucl-ex/0410020}{{\ttfamily arXiv:nucl-ex/0410020
  [nucl-ex]}}.

\bibitem{Back:2004je}
{\bfseries PHOBOS} Collaboration, B.~B. Back {\em et~al.}, ``{The PHOBOS
  perspective on discoveries at RHIC}'',
  \href{http://dx.doi.org/10.1016/j.nuclphysa.2005.03.084}{{\em Nucl. Phys.}
  {\bfseries A757} (2005) 28--101},
\href{http://arxiv.org/abs/nucl-ex/0410022}{{\ttfamily arXiv:nucl-ex/0410022
  [nucl-ex]}}.

\bibitem{Aamodt:2010pa}
{\bfseries ALICE} Collaboration, K.~Aamodt {\em et~al.}, ``{Elliptic flow of
  charged particles in Pb--Pb collisions at 2.76 TeV}'',
  \href{http://dx.doi.org/10.1103/PhysRevLett.105.252302}{{\em Phys. Rev.
  Lett.} {\bfseries 105} (2010) 252302},
\href{http://arxiv.org/abs/1011.3914}{{\ttfamily arXiv:1011.3914 [nucl-ex]}}.

\bibitem{Aamodt:2010jd}
{\bfseries ALICE} Collaboration, K.~Aamodt {\em et~al.}, ``{Suppression of
  Charged Particle Production at Large Transverse Momentum in Central Pb--Pb
  Collisions at $\sqrt{s_{\mathrm{NN}}} =$ 2.76 TeV}'',
  \href{http://dx.doi.org/10.1016/j.physletb.2010.12.020}{{\em Phys. Lett.}
  {\bfseries B696} (2011) 30--39},
\href{http://arxiv.org/abs/1012.1004}{{\ttfamily arXiv:1012.1004 [nucl-ex]}}.

\bibitem{Schukraft:2011na}
{\bfseries ALICE} Collaboration, J.~Schukraft, ``{Heavy Ion physics with the
  ALICE experiment at the CERN LHC}'',
  \href{http://dx.doi.org/10.1098/rsta.2011.0469}{{\em Phil. Trans. Roy. Soc.
  Lond.} {\bfseries A370} (2012) 917--932},
\href{http://arxiv.org/abs/1109.4291}{{\ttfamily arXiv:1109.4291 [hep-ex]}}.

\bibitem{Cabibbo:1975ig}
N.~Cabibbo and G.~Parisi, ``{Exponential Hadronic Spectrum and Quark
  Liberation}'',
\href{http://dx.doi.org/10.1016/0370-2693(75)90158-6}{{\em Phys. Lett.}
  {\bfseries 59B} (1975) 67--69}.

\bibitem{Shuryak:1978ij}
E.~V. Shuryak, ``{Quark-Gluon Plasma and Hadronic Production of Leptons,
  Photons and Psions}'',
  \href{http://dx.doi.org/10.1016/0370-2693(78)90370-2}{{\em Phys. Lett.}
  {\bfseries 78B} (1978) 150}.
[Yad. Fiz.28,796(1978)].

\bibitem{Laermann:2003cv}
E.~Laermann and O.~Philipsen, ``{The Status of lattice QCD at finite
  temperature}'',
  \href{http://dx.doi.org/10.1146/annurev.nucl.53.041002.110609}{{\em Ann. Rev.
  Nucl. Part. Sci.} {\bfseries 53} (2003) 163--198},
\href{http://arxiv.org/abs/hep-ph/0303042}{{\ttfamily arXiv:hep-ph/0303042
  [hep-ph]}}.

\bibitem{Aoki:2006we}
Y.~Aoki, G.~Endrodi, Z.~Fodor, S.~D. Katz, and K.~K. Szabo, ``{The Order of the
  quantum chromodynamics transition predicted by the standard model of particle
  physics}'', \href{http://dx.doi.org/10.1038/nature05120}{{\em Nature}
  {\bfseries 443} (2006) 675--678},
\href{http://arxiv.org/abs/hep-lat/0611014}{{\ttfamily arXiv:hep-lat/0611014
  [hep-lat]}}.

\bibitem{Gupta:2011wh}
S.~Gupta, X.~Luo, B.~Mohanty, H.~G. Ritter, and N.~Xu, ``{Scale for the Phase
  Diagram of Quantum Chromodynamics}'',
  \href{http://dx.doi.org/10.1126/science.1204621}{{\em Science} {\bfseries
  332} (2011) 1525--1528},
\href{http://arxiv.org/abs/1105.3934}{{\ttfamily arXiv:1105.3934 [hep-ph]}}.

\bibitem{ALICE:2012xs}
{\bfseries ALICE} Collaboration, B.~Abelev {\em et~al.}, ``{Pseudorapidity
  density of charged particles in $p$ + Pb collisions at
  $\sqrt{s_{\mathrm{NN}}}=5.02$ TeV}'',
  \href{http://dx.doi.org/10.1103/PhysRevLett.110.032301}{{\em Phys. Rev.
  Lett.} {\bfseries 110} no.~3, (2013) 032301},
\href{http://arxiv.org/abs/1210.3615}{{\ttfamily arXiv:1210.3615 [nucl-ex]}}.

\bibitem{Adam:2014qja}
{\bfseries ALICE} Collaboration, J.~Adam {\em et~al.}, ``{Centrality dependence
  of particle production in p--Pb collisions at $\sqrt{s_{\rm NN} }$= 5.02
  TeV}'', \href{http://dx.doi.org/10.1103/PhysRevC.91.064905}{{\em Phys. Rev.}
  {\bfseries C91} no.~6, (2015) 064905},
\href{http://arxiv.org/abs/1412.6828}{{\ttfamily arXiv:1412.6828 [nucl-ex]}}.

\bibitem{Aamodt:2010cz}
{\bfseries ALICE} Collaboration, K.~Aamodt {\em et~al.}, ``{Centrality
  dependence of the charged-particle multiplicity density at mid-rapidity in
  Pb--Pb collisions at $\sqrt{s_{\mathrm{NN}}}=2.76$ TeV}'',
  \href{http://dx.doi.org/10.1103/PhysRevLett.106.032301}{{\em Phys. Rev.
  Lett.} {\bfseries 106} (2011) 032301},
\href{http://arxiv.org/abs/1012.1657}{{\ttfamily arXiv:1012.1657 [nucl-ex]}}.

\bibitem{Arneodo:1992wf}
M.~Arneodo, ``{Nuclear effects in structure functions}'',
\href{http://dx.doi.org/10.1016/0370-1573(94)90048-5}{{\em Phys. Rept.}
  {\bfseries 240} (1994) 301--393}.

\bibitem{Abelev:2014hha}
{\bfseries ALICE} Collaboration, B.~Abelev {\em et~al.}, ``{Measurement of
  prompt $D$-meson production in p--Pb collisions at $\sqrt{s_{\mathrm{NN}}}$ =
  5.02 TeV}'', \href{http://dx.doi.org/10.1103/PhysRevLett.113.232301}{{\em
  Phys. Rev. Lett.} {\bfseries 113} no.~23, (2014) 232301},
\href{http://arxiv.org/abs/1405.3452}{{\ttfamily arXiv:1405.3452 [nucl-ex]}}.

\bibitem{Alver:2010ck}
{\bfseries PHOBOS} Collaboration, B.~Alver {\em et~al.}, ``{Phobos results on
  charged particle multiplicity and pseudorapidity distributions in Au+Au,
  Cu+Cu, d+Au, and p+p collisions at ultra-relativistic energies}'',
  \href{http://dx.doi.org/10.1103/PhysRevC.83.024913}{{\em Phys. Rev.}
  {\bfseries C83} (2011) 024913},
\href{http://arxiv.org/abs/1011.1940}{{\ttfamily arXiv:1011.1940 [nucl-ex]}}.

\bibitem{ALICE:2017jyt}
{\bfseries ALICE} Collaboration, J.~Adam {\em et~al.}, ``{Enhanced production
  of multi-strange hadrons in high-multiplicity proton-proton collisions}'',
  \href{http://dx.doi.org/10.1038/nphys4111}{{\em Nature Phys.} {\bfseries 13}
  (2017) 535--539},
\href{http://arxiv.org/abs/1606.07424}{{\ttfamily arXiv:1606.07424 [nucl-ex]}}.

\bibitem{Aggarwal:2010mt}
{\bfseries STAR} Collaboration, M.~M. Aggarwal {\em et~al.}, ``{K$^{*0}$
  production in Cu+Cu and Au+Au collisions at $\sqrt{s_{\mathrm{NN}}}$ = 62.4
  GeV and 200 GeV}'', \href{http://dx.doi.org/10.1103/PhysRevC.84.034909}{{\em
  Phys. Rev.} {\bfseries C84} (2011) 034909},
\href{http://arxiv.org/abs/1006.1961}{{\ttfamily arXiv:1006.1961 [nucl-ex]}}.

\bibitem{Abelev:2014uua}
{\bfseries ALICE} Collaboration, B.~Abelev {\em et~al.}, ``{K$^{*}$(892)$^{0}$
  and $\phi$(1020) production in Pb--Pb collisions at $\sqrt{s_{\mathrm{NN}}}$
  = 2.76 TeV}'', \href{http://dx.doi.org/10.1103/PhysRevC.91.024609}{{\em Phys.
  Rev.} {\bfseries C91} (2015) 024609},
\href{http://arxiv.org/abs/1404.0495}{{\ttfamily arXiv:1404.0495 [nucl-ex]}}.

\bibitem{Adam:2017zbf}
{\bfseries ALICE} Collaboration, J.~Adam {\em et~al.}, ``{K$^{*}$(892)$^{0}$
  and $\phi$(1020) meson production at high transverse momentum in pp and
  Pb--Pb collisions at $\sqrt{s_\mathrm{NN}}$ = 2.76 TeV}'',
  \href{http://dx.doi.org/10.1103/PhysRevC.95.064606}{{\em Phys. Rev.}
  {\bfseries C95} no.~6, (2017) 064606},
\href{http://arxiv.org/abs/1702.00555}{{\ttfamily arXiv:1702.00555 [nucl-ex]}}.

\bibitem{Adams:2006yu}
{\bfseries STAR} Collaboration, B.~I. Abelev {\em et~al.}, ``{Strange baryon
  resonance production in \mbox{$\sqrt{s_{\rm NN}}$ = 200 GeV} p+p and Au+Au
  collisions}'', \href{http://dx.doi.org/10.1103/PhysRevLett.97.132301}{{\em
  Phys. Rev. Lett.} {\bfseries 97} (2006) 132301},
\href{http://arxiv.org/abs/nucl-ex/0604019}{{\ttfamily arXiv:nucl-ex/0604019
  [nucl-ex]}}.

\bibitem{Adam:2016bpr}
{\bfseries ALICE} Collaboration, J.~Adam {\em et~al.}, ``{Production of
  K$^{*}$(892)$^{0}$ and $\phi$(1020) in p–Pb collisions at
  $\sqrt{s_{{\mathrm{NN}}}}$ = 5.02 TeV}'',
  \href{http://dx.doi.org/10.1140/epjc/s10052-016-4088-7}{{\em Eur. Phys. J.}
  {\bfseries C76} no.~5, (2016) 245},
\href{http://arxiv.org/abs/1601.07868}{{\ttfamily arXiv:1601.07868 [nucl-ex]}}.

\bibitem{Stachel:2013zma}
J.~Stachel, A.~Andronic, P.~Braun-Munzinger, and K.~Redlich, ``{Confronting LHC
  data with the statistical hadronization model}'',
  \href{http://dx.doi.org/10.1088/1742-6596/509/1/012019}{{\em J. Phys. Conf.
  Ser.} {\bfseries 509} (2014) 012019},
\href{http://arxiv.org/abs/1311.4662}{{\ttfamily arXiv:1311.4662 [nucl-th]}}.

\bibitem{Vovchenko:2019kes}
V.~Vovchenko, B.~Dönigus, and H.~Stoecker, ``{Canonical statistical model
  analysis of p-p, p-Pb, and Pb-Pb collisions at the LHC}'',
\href{http://arxiv.org/abs/1906.03145}{{\ttfamily arXiv:1906.03145 [hep-ph]}}.

\bibitem{Acharya:2018qnp}
{\bfseries ALICE} Collaboration, S.~Acharya {\em et~al.}, ``{Production of the
  $\rho$(770)${^{0}}$ meson in pp and Pb--Pb collisions at $\sqrt{s_{\rm NN}}$
  = 2.76 TeV}'', \href{http://dx.doi.org/10.1103/PhysRevC.99.064901}{{\em Phys.
  Rev.} {\bfseries C99} no.~6, (2019) 064901},
\href{http://arxiv.org/abs/1805.04365}{{\ttfamily arXiv:1805.04365 [nucl-ex]}}.

\bibitem{Tanabashi:2018oca}
{\bfseries Particle Data Group} Collaboration, M.~Tanabashi {\em et~al.},
  ``{Review of Particle Physics}'',
\href{http://dx.doi.org/10.1103/PhysRevD.98.030001}{{\em Phys. Rev.} {\bfseries
  D98} no.~3, (2018) 030001}.

\bibitem{Abelev:2008yz}
{\bfseries STAR} Collaboration, B.~I. Abelev {\em et~al.}, ``{Hadronic
  resonance production in d+Au collisions at $\sqrt{s_{\rm NN}}$ = 200 GeV at
  RHIC}'', \href{http://dx.doi.org/10.1103/PhysRevC.78.044906}{{\em Phys. Rev.}
  {\bfseries C78} (2008) 044906},
\href{http://arxiv.org/abs/0801.0450}{{\ttfamily arXiv:0801.0450 [nucl-ex]}}.

\bibitem{ALICE:2018ewo}
{\bfseries ALICE} Collaboration, S.~Acharya {\em et~al.}, ``{Suppression of
  $\Lambda(1520)$ resonance production in central Pb--Pb collisions at
  $\sqrt{s_{\rm NN}}$ = 2.76 TeV}'',
  \href{http://dx.doi.org/10.1103/PhysRevC.99.024905}{{\em Phys. Rev.}
  {\bfseries C99} (2019) 024905},
\href{http://arxiv.org/abs/1805.04361}{{\ttfamily arXiv:1805.04361 [nucl-ex]}}.

\bibitem{Drescher:2000ha}
H.~J. Drescher, M.~Hladik, S.~Ostapchenko, T.~Pierog, and K.~Werner, ``{Parton
  based Gribov-Regge theory}'',
  \href{http://dx.doi.org/10.1016/S0370-1573(00)00122-8}{{\em Phys. Rept.}
  {\bfseries 350} (2001) 93--289},
\href{http://arxiv.org/abs/hep-ph/0007198}{{\ttfamily arXiv:hep-ph/0007198
  [hep-ph]}}.

\bibitem{Werner:2010aa}
K.~Werner, I.~Karpenko, T.~Pierog, M.~Bleicher, and K.~Mikhailov,
  ``{Event-by-Event Simulation of the Three-Dimensional Hydrodynamic Evolution
  from Flux Tube Initial Conditions in Ultrarelativistic Heavy Ion
  Collisions}'', \href{http://dx.doi.org/10.1103/PhysRevC.82.044904}{{\em Phys.
  Rev.} {\bfseries C82} (2010) 044904},
\href{http://arxiv.org/abs/1004.0805}{{\ttfamily arXiv:1004.0805 [nucl-th]}}.

\bibitem{Werner:2013tya}
K.~Werner, B.~Guiot, I.~Karpenko, and T.~Pierog, ``{Analysing radial flow
  features in p--Pb and pp collisions at several TeV by studying identified
  particle production in EPOS3}'',
  \href{http://dx.doi.org/10.1103/PhysRevC.89.064903}{{\em Phys. Rev.}
  {\bfseries C89} no.~6, (2014) 064903},
\href{http://arxiv.org/abs/1312.1233}{{\ttfamily arXiv:1312.1233 [nucl-th]}}.

\bibitem{Bass:1998ca}
S.~A. Bass {\em et~al.}, ``{Microscopic models for ultrarelativistic heavy ion
  collisions}'', \href{http://dx.doi.org/10.1016/S0146-6410(98)00058-1}{{\em
  Prog. Part. Nucl. Phys.} {\bfseries 41} (1998) 255--369},
  \href{http://arxiv.org/abs/nucl-th/9803035}{{\ttfamily arXiv:nucl-th/9803035
  [nucl-th]}}.
[Prog. Part. Nucl. Phys.41,225(1998)].

\bibitem{Knospe:2015nva}
A.~G. Knospe, C.~Markert, K.~Werner, J.~Steinheimer, and M.~Bleicher,
  ``{Hadronic resonance production and interaction in partonic and hadronic
  matter in the EPOS3 model with and without the hadronic afterburner UrQMD}'',
  \href{http://dx.doi.org/10.1103/PhysRevC.93.014911}{{\em Phys. Rev.}
  {\bfseries C93} no.~1, (2016) 014911},
\href{http://arxiv.org/abs/1509.07895}{{\ttfamily arXiv:1509.07895 [nucl-th]}}.

\bibitem{Abelev:2012ola}
{\bfseries ALICE} Collaboration, B.~Abelev {\em et~al.}, ``{Long-range angular
  correlations on the near and away side in $p$-Pb collisions at
  $\sqrt{s_{\mathrm{NN}}}=5.02$ TeV}'',
  \href{http://dx.doi.org/10.1016/j.physletb.2013.01.012}{{\em Phys. Lett.}
  {\bfseries B719} (2013) 29--41},
\href{http://arxiv.org/abs/1212.2001}{{\ttfamily arXiv:1212.2001 [nucl-ex]}}.

\bibitem{Abelev:2013wsa}
{\bfseries ALICE} Collaboration, B.~Abelev {\em et~al.}, ``{Long-range angular
  correlations of $\rm \pi$, K and p in p--Pb collisions at $\sqrt{s_{\rm NN}}$
  = 5.02 TeV}'', \href{http://dx.doi.org/10.1016/j.physletb.2013.08.024}{{\em
  Phys. Lett.} {\bfseries B726} (2013) 164--177},
\href{http://arxiv.org/abs/1307.3237}{{\ttfamily arXiv:1307.3237 [nucl-ex]}}.

\bibitem{Abelev:2014mda}
{\bfseries ALICE} Collaboration, B.~Abelev {\em et~al.}, ``{Multiparticle
  azimuthal correlations in p--Pb and Pb--Pb collisions at the CERN Large
  Hadron Collider}'', \href{http://dx.doi.org/10.1103/PhysRevC.90.054901}{{\em
  Phys. Rev.} {\bfseries C90} no.~5, (2014) 054901},
\href{http://arxiv.org/abs/1406.2474}{{\ttfamily arXiv:1406.2474 [nucl-ex]}}.

\bibitem{Bozek:2011if}
P.~Bozek, ``{Collective flow in p--Pb and d--Pd collisions at TeV energies}'',
  \href{http://dx.doi.org/10.1103/PhysRevC.85.014911}{{\em Phys. Rev.}
  {\bfseries C85} (2012) 014911},
\href{http://arxiv.org/abs/1112.0915}{{\ttfamily arXiv:1112.0915 [hep-ph]}}.

\bibitem{Bozek:2012gr}
P.~Bozek and W.~Broniowski, ``{Correlations from hydrodynamic flow in p--Pb
  collisions}'', \href{http://dx.doi.org/10.1016/j.physletb.2012.12.051}{{\em
  Phys. Lett.} {\bfseries B718} (2013) 1557--1561},
\href{http://arxiv.org/abs/1211.0845}{{\ttfamily arXiv:1211.0845 [nucl-th]}}.

\bibitem{Abelev:2013haa}
{\bfseries ALICE} Collaboration, B.~Abelev {\em et~al.}, ``{Multiplicity
  Dependence of Pion, Kaon, Proton and Lambda Production in p--Pb Collisions at
  $\sqrt{s_{\mathrm{NN}}}$ = 5.02 TeV}'',
  \href{http://dx.doi.org/10.1016/j.physletb.2013.11.020}{{\em Phys. Lett.}
  {\bfseries B728} (2014) 25--38},
\href{http://arxiv.org/abs/1307.6796}{{\ttfamily arXiv:1307.6796 [nucl-ex]}}.

\bibitem{Adam:2015vsf}
{\bfseries ALICE} Collaboration, J.~Adam {\em et~al.}, ``{Multi-strange baryon
  production in p--Pb collisions at $\sqrt{s_\mathrm{NN}}$ = 5.02 TeV}'',
  \href{http://dx.doi.org/10.1016/j.physletb.2016.05.027}{{\em Phys. Lett.}
  {\bfseries B758} (2016) 389--401},
\href{http://arxiv.org/abs/1512.07227}{{\ttfamily arXiv:1512.07227 [nucl-ex]}}.

\bibitem{Aamodt:2008zz}
{\bfseries ALICE} Collaboration, K.~Aamodt {\em et~al.}, ``{The ALICE
  experiment at the CERN LHC}'',
\href{http://dx.doi.org/10.1088/1748-0221/3/08/S08002}{{\em JINST} {\bfseries
  3} (2008) S08002}.

\bibitem{Abelev:2014ffa}
{\bfseries ALICE} Collaboration, B.~Abelev {\em et~al.}, ``{Performance of the
  ALICE Experiment at the CERN LHC}'',
  \href{http://dx.doi.org/10.1142/S0217751X14300440}{{\em Int. J. Mod. Phys.}
  {\bfseries A29} (2014) 1430044},
\href{http://arxiv.org/abs/1402.4476}{{\ttfamily arXiv:1402.4476 [nucl-ex]}}.

\bibitem{Aamodt:2010aa}
{\bfseries ALICE} Collaboration, K.~Aamodt {\em et~al.}, ``{Alignment of the
  ALICE Inner Tracking System with cosmic-ray tracks}'',
  \href{http://dx.doi.org/10.1088/1748-0221/5/03/P03003}{{\em JINST} {\bfseries
  5} (2010) P03003},
\href{http://arxiv.org/abs/1001.0502}{{\ttfamily arXiv:1001.0502
  [physics.ins-det]}}.

\bibitem{Alme:2010ke}
J.~Alme {\em et~al.}, ``{The ALICE TPC, a large 3-dimensional tracking device
  with fast readout for ultra-high multiplicity events}'',
  \href{http://dx.doi.org/10.1016/j.nima.2010.04.042}{{\em Nucl. Instrum.
  Meth.} {\bfseries A622} (2010) 316--367},
\href{http://arxiv.org/abs/1001.1950}{{\ttfamily arXiv:1001.1950
  [physics.ins-det]}}.

\bibitem{Akindinov:2013tea}
A.~Akindinov {\em et~al.}, ``{Performance of the ALICE Time-Of-Flight detector
  at the LHC}'',
\href{http://dx.doi.org/10.1140/epjp/i2013-13044-x}{{\em Eur. Phys. J. Plus}
  {\bfseries 128} (2013) 44}.

\bibitem{Abbas:2013taa}
{\bfseries ALICE} Collaboration, E.~Abbas {\em et~al.}, ``{Performance of the
  ALICE VZERO system}'',
  \href{http://dx.doi.org/10.1088/1748-0221/8/10/P10016}{{\em JINST} {\bfseries
  8} (2013) P10016},
\href{http://arxiv.org/abs/1306.3130}{{\ttfamily arXiv:1306.3130 [nucl-ex]}}.

\bibitem{Cortese:781854}
{\bfseries ALICE} Collaboration, P.~Cortese {\em et~al.}, {\em {ALICE forward
  detectors: FMD, TO and VO: Technical Design Report}}.
\newblock Technical Design Report ALICE. CERN, Geneva, 2004.
\newblock \url{http://cds.cern.ch/record/781854}.
\newblock Submitted on 10 Sep 2004.

\bibitem{Gallio:381433}
{\bfseries ALICE} Collaboration, M.~Gallio, W.~Klempt, L.~Leistam, J.~De~Groot,
  and J.~Schükraft, {\em {ALICE Zero-Degree Calorimeter (ZDC): Technical
  Design Report}}.
\newblock Technical Design Report ALICE. CERN, Geneva, 1999.
\newblock \url{https://cds.cern.ch/record/381433}.

\bibitem{Adam:2015qaa}
{\bfseries ALICE} Collaboration, J.~Adam {\em et~al.}, ``{Measurement of pion,
  kaon and proton production in proton-proton collisions at $\sqrt{s} = 7$
  TeV}'', \href{http://dx.doi.org/10.1140/epjc/s10052-015-3422-9}{{\em Eur.
  Phys. J.} {\bfseries C75} no.~5, (2015) 226},
\href{http://arxiv.org/abs/1504.00024}{{\ttfamily arXiv:1504.00024 [nucl-ex]}}.

\bibitem{Abelev:2014oea}
{\bfseries ALICE} Collaboration, B.~Abelev {\em et~al.}, ``{Production of
  inclusive $\Upsilon$(1S) and $\Upsilon$(2S) in p--Pb collisions at
  $\mathbf{\sqrt{s_{{\rm NN}}} = 5.02}$ TeV}'',
  \href{http://dx.doi.org/10.1016/j.physletb.2014.11.041}{{\em Phys. Lett.}
  {\bfseries B740} (2015) 105--117},
\href{http://arxiv.org/abs/1410.2234}{{\ttfamily arXiv:1410.2234 [nucl-ex]}}.

\bibitem{Adam:2015gka}
{\bfseries ALICE} Collaboration, J.~Adam {\em et~al.}, ``{Charged-particle
  multiplicities in proton-proton collisions at $\sqrt{s}$ = 0.9 to 8 TeV}'',
  \href{http://dx.doi.org/10.1140/epjc/s10052-016-4571-1}{{\em Eur. Phys. J.}
  {\bfseries C77} no.~1, (2017) 33},
\href{http://arxiv.org/abs/1509.07541}{{\ttfamily arXiv:1509.07541 [nucl-ex]}}.

\bibitem{Aamodt:2011zj}
{\bfseries ALICE} Collaboration, K.~Aamodt {\em et~al.}, ``{Production of
  pions, kaons and protons in $pp$ collisions at $\sqrt{s}$ = 900 GeV with
  ALICE at the LHC}'',
  \href{http://dx.doi.org/10.1140/epjc/s10052-011-1655-9}{{\em Eur. Phys. J.}
  {\bfseries C71} (2011) 1655},
\href{http://arxiv.org/abs/1101.4110}{{\ttfamily arXiv:1101.4110 [hep-ex]}}.

\bibitem{Abelev:2012hy}
{\bfseries ALICE} Collaboration, B.~Abelev {\em et~al.}, ``{Production of
  $K^*(892)^0$ and $\phi(1020)$ in $pp$ collisions at $\sqrt{s}=7$ TeV}'',
  \href{http://dx.doi.org/10.1140/epjc/s10052-012-2183-y}{{\em Eur. Phys. J.}
  {\bfseries C72} (2012) 2183},
\href{http://arxiv.org/abs/1208.5717}{{\ttfamily arXiv:1208.5717 [hep-ex]}}.

\bibitem{Skands:2010ak}
P.~Z. Skands, ``{Tuning Monte Carlo Generators: The Perugia Tunes}'',
  \href{http://dx.doi.org/10.1103/PhysRevD.82.074018}{{\em Phys. Rev.}
  {\bfseries D82} (2010) 074018},
\href{http://arxiv.org/abs/1005.3457}{{\ttfamily arXiv:1005.3457 [hep-ph]}}.

\bibitem{Roesler:2000he}
S.~Roesler, R.~Engel, and J.~Ranft,
  \href{http://dx.doi.org/10.1007/978-3-642-18211-2_166}{``{The Monte Carlo
  event generator DPMJET-III}'',} in {\em {Advanced Monte Carlo for radiation
  physics, particle transport simulation and applications. Proceedings,
  Conference, MC2000, Lisbon, Portugal, October 23-26, 2000}}, pp.~1033--1038.
\newblock 2000.
\newblock \href{http://arxiv.org/abs/hep-ph/0012252}{{\ttfamily
  arXiv:hep-ph/0012252 [hep-ph]}}.
\newblock
\url{http://www-public.slac.stanford.edu/sciDoc/docMeta.aspx?slacPubNumber=SLAC-PUB-8740}.
\newblock

\bibitem{Brun:1994aa}
R.~Brun, F.~Bruyant, F.~Carminati, S.~Giani, M.~Maire, A.~McPherson,
  G.~Patrick, and L.~Urban,
``{GEANT Detector Description and Simulation Tool}'',.

\bibitem{Abelev:2013vea}
{\bfseries ALICE} Collaboration, B.~Abelev {\em et~al.}, ``{Centrality
  dependence of $\pi$, K, p production in Pb--Pb collisions at
  $\sqrt{s_{\mathrm{NN}}}$ = 2.76 TeV}'',
  \href{http://dx.doi.org/10.1103/PhysRevC.88.044910}{{\em Phys. Rev.}
  {\bfseries C88} (2013) 044910},
\href{http://arxiv.org/abs/1303.0737}{{\ttfamily arXiv:1303.0737 [hep-ex]}}.

\bibitem{Battistoni:2007zzb}
G.~Battistoni, S.~Muraro, P.~R. Sala, F.~Cerutti, A.~Ferrari, S.~Roesler,
  A.~Fasso, and J.~Ranft, ``{The FLUKA code: Description and benchmarking}'',
\href{http://dx.doi.org/10.1063/1.2720455}{{\em AIP Conf. Proc.} {\bfseries
  896} no.~1, (2007) 31--49}.

\bibitem{Tsallis:1987eu}
C.~Tsallis, ``{Possible Generalization of Boltzmann-Gibbs Statistics}'',
\href{http://dx.doi.org/10.1007/BF01016429}{{\em J. Statist. Phys.} {\bfseries
  52} (1988) 479--487}.

\bibitem{Schnedermann:1993ws}
E.~Schnedermann, J.~Sollfrank, and U.~W. Heinz, ``{Thermal phenomenology of
  hadrons from \mbox{200 A/GeV} S+S collisions}'',
  \href{http://dx.doi.org/10.1103/PhysRevC.48.2462}{{\em Phys. Rev.} {\bfseries
  C48} (1993) 2462--2475},
\href{http://arxiv.org/abs/nucl-th/9307020}{{\ttfamily arXiv:nucl-th/9307020
  [nucl-th]}}.

\bibitem{Abelev:2008ab}
{\bfseries STAR} Collaboration, B.~I. Abelev {\em et~al.}, ``{Systematic
  Measurements of Identified Particle Spectra in pp, d+Au and Au+Au Collisions
  from STAR}'', \href{http://dx.doi.org/10.1103/PhysRevC.79.034909}{{\em Phys.
  Rev.} {\bfseries C79} (2009) 034909},
\href{http://arxiv.org/abs/0808.2041}{{\ttfamily arXiv:0808.2041 [nucl-ex]}}.

\bibitem{Adamova:2017elh}
{\bfseries ALICE} Collaboration, D.~Adamova {\em et~al.}, ``{Production of
  $\Sigma(1385)^{\pm}$ and $\Xi(1530)^{0}$ in p--Pb collisions at $\sqrt{s_{\rm
  NN}}=5.02$ TeV}'',
  \href{http://dx.doi.org/10.1140/epjc/s10052-017-4943-1}{{\em Eur. Phys. J.}
  {\bfseries C77} no.~6, (2017) 389},
\href{http://arxiv.org/abs/1701.07797}{{\ttfamily arXiv:1701.07797 [nucl-ex]}}.

\bibitem{Abelev:2014qqa}
{\bfseries ALICE} Collaboration, B.~Abelev {\em et~al.}, ``{Production of
  $\Sigma(1385)^{\pm}$ and $\Xi(1530)^{0}$ in proton-proton collisions at
  $\sqrt{s}=$ 7 TeV}'',
  \href{http://dx.doi.org/10.1140/epjc/s10052-014-3191-x}{{\em Eur. Phys. J.}
  {\bfseries C75} no.~1, (2015) 1},
\href{http://arxiv.org/abs/1406.3206}{{\ttfamily arXiv:1406.3206 [nucl-ex]}}.

\bibitem{Heinz:2004qz}
U.~W. Heinz, ``{Concepts of heavy ion physics}'', in {\em {2002 European School
  of high-energy physics, Pylos, Greece, 25 Aug-7 Sep 2002: Proceedings}},
  pp.~165--238.
\newblock 2004.
\newblock \href{http://arxiv.org/abs/hep-ph/0407360}{{\ttfamily
  arXiv:hep-ph/0407360 [hep-ph]}}.
\newblock
\url{http://doc.cern.ch/yellowrep/CERN-2004-001}.
\newblock

\bibitem{Sollfrank:1997bq}
J.~Sollfrank, F.~Becattini, K.~Redlich, and H.~Satz, ``{Canonical strangeness
  enhancement}'', \href{http://dx.doi.org/10.1016/S0375-9474(98)00395-9}{{\em
  Nucl. Phys.} {\bfseries A638} (1998) 399C--402C},
\href{http://arxiv.org/abs/nucl-th/9802046}{{\ttfamily arXiv:nucl-th/9802046
  [nucl-th]}}.

\bibitem{Letessier:2006wn}
J.~Letessier and J.~Rafelski, ``{Strangeness chemical equilibration in QGP at
  RHIC and CERN LHC}'',
  \href{http://dx.doi.org/10.1103/PhysRevC.75.014905}{{\em Phys. Rev.}
  {\bfseries C75} (2007) 014905},
\href{http://arxiv.org/abs/nucl-th/0602047}{{\ttfamily arXiv:nucl-th/0602047
  [nucl-th]}}.

\bibitem{Abelev:2006cs}
{\bfseries STAR} Collaboration, B.~I. Abelev {\em et~al.}, ``{Strange particle
  production in p+p collisions at \mbox{$\sqrt{s}$ = 200 GeV}}'',
  \href{http://dx.doi.org/10.1103/PhysRevC.75.064901}{{\em Phys. Rev.}
  {\bfseries C75} (2007) 064901},
\href{http://arxiv.org/abs/nucl-ex/0607033}{{\ttfamily arXiv:nucl-ex/0607033
  [nucl-ex]}}.

\bibitem{Wheaton:2004vg}
S.~Wheaton and J.~Cleymans, ``{Statistical-thermal model calculations using
  THERMUS}'', \href{http://dx.doi.org/10.1088/0954-3899/31/6/060}{{\em J.
  Phys.} {\bfseries G31} (2005) S1069--S1074},
\href{http://arxiv.org/abs/hep-ph/0412031}{{\ttfamily arXiv:hep-ph/0412031
  [hep-ph]}}.

\bibitem{Andronic:2005yp}
A.~Andronic, P.~Braun-Munzinger, and J.~Stachel, ``{Hadron production in
  central nucleus-nucleus collisions at chemical freeze-out}'',
  \href{http://dx.doi.org/10.1016/j.nuclphysa.2006.03.012}{{\em Nucl. Phys.}
  {\bfseries A772} (2006) 167--199},
\href{http://arxiv.org/abs/nucl-th/0511071}{{\ttfamily arXiv:nucl-th/0511071
  [nucl-th]}}.

\bibitem{Petran:2013dva}
M.~Petran, J.~Letessier, J.~Rafelski, and G.~Torrieri, ``{SHARE with CHARM}'',
  \href{http://dx.doi.org/10.1016/j.cpc.2014.02.026}{{\em Comput. Phys.
  Commun.} {\bfseries 185} (2014) 2056--2079},
\href{http://arxiv.org/abs/1310.5108}{{\ttfamily arXiv:1310.5108 [hep-ph]}}.

\end{thebibliography}\endgroup
    
    \newpage
    \appendix
    
    \section{The ALICE Collaboration}
    \label{app:collab}

\begingroup
\small
\begin{flushleft}
S.~Acharya\Irefn{org141}\And 
D.~Adamov\'{a}\Irefn{org93}\And 
S.P.~Adhya\Irefn{org141}\And 
A.~Adler\Irefn{org73}\And 
J.~Adolfsson\Irefn{org79}\And 
M.M.~Aggarwal\Irefn{org98}\And 
G.~Aglieri Rinella\Irefn{org34}\And 
M.~Agnello\Irefn{org31}\And 
N.~Agrawal\Irefn{org10}\textsuperscript{,}\Irefn{org48}\textsuperscript{,}\Irefn{org53}\And 
Z.~Ahammed\Irefn{org141}
% ...
\And
\renewcommand\labelenumi{\textsuperscript{\theenumi}~}

\section*{Affiliation notes}
\renewcommand\theenumi{\roman{enumi}}
\begin{Authlist}
\item \Adef{org*}Deceased
\item \Adef{orgI}Dipartimento DET del Politecnico di Torino, Turin, Italy
\item \Adef{orgII}M.V. Lomonosov Moscow State University, D.V. Skobeltsyn Institute of Nuclear, Physics, Moscow, Russia
\item \Adef{orgIII}Department of Applied Physics, Aligarh Muslim University, Aligarh, India
\item \Adef{orgIV}Institute of Theoretical Physics, University of Wroclaw, Poland
\end{Authlist}

\section*{Collaboration Institutes}
\renewcommand\theenumi{\arabic{enumi}~}
\begin{Authlist}
\item \Idef{org1}A.I. Alikhanyan National Science Laboratory (Yerevan Physics Institute) Foundation, Yerevan, Armenia
\item \Idef{org2}Bogolyubov Institute for Theoretical Physics, National Academy of Sciences of Ukraine, Kiev, Ukraine
\item \Idef{org3}Bose Institute, Department of Physics  and Centre for Astroparticle Physics and Space Science (CAPSS), Kolkata, India
\item \Idef{org4}Budker Institute for Nuclear Physics, Novosibirsk, Russia
\item \Idef{org5}California Polytechnic State University, San Luis Obispo, California, United States
\item \Idef{org6}Central China Normal University, Wuhan, China
\item \Idef{org7}Centre de Calcul de l'IN2P3, Villeurbanne, Lyon, France
\item \Idef{org8}Centro de Aplicaciones Tecnol\'{o}gicas y Desarrollo Nuclear (CEADEN), Havana, Cuba
\item \Idef{org9}Centro de Investigaci\'{o}n y de Estudios Avanzados (CINVESTAV), Mexico City and M\'{e}rida, Mexico
\item \Idef{org10}Centro Fermi - Museo Storico della Fisica e Centro Studi e Ricerche ``Enrico Fermi', Rome, Italy
\item \Idef{org11}Chicago State University, Chicago, Illinois, United States
\item \Idef{org12}China Institute of Atomic Energy, Beijing, China
\item \Idef{org13}Chonbuk National University, Jeonju, Republic of Korea
\item \Idef{org14}Comenius University Bratislava, Faculty of Mathematics, Physics and Informatics, Bratislava, Slovakia
\item \Idef{org15}COMSATS University Islamabad, Islamabad, Pakistan
\item \Idef{org16}Creighton University, Omaha, Nebraska, United States
\item \Idef{org17}Department of Physics, Aligarh Muslim University, Aligarh, India
\item \Idef{org18}Department of Physics, Pusan National University, Pusan, Republic of Korea
\item \Idef{org19}Department of Physics, Sejong University, Seoul, Republic of Korea
\item \Idef{org20}Department of Physics, University of California, Berkeley, California, United States
\item \Idef{org21}Department of Physics, University of Oslo, Oslo, Norway
\item \Idef{org22}Department of Physics and Technology, University of Bergen, Bergen, Norway
\item \Idef{org23}Dipartimento di Fisica dell'Universit\`{a} 'La Sapienza' and Sezione INFN, Rome, Italy
\item \Idef{org24}Dipartimento di Fisica dell'Universit\`{a} and Sezione INFN, Cagliari, Italy
\item \Idef{org25}Dipartimento di Fisica dell'Universit\`{a} and Sezione INFN, Trieste, Italy
\item \Idef{org26}Dipartimento di Fisica dell'Universit\`{a} and Sezione INFN, Turin, Italy
\item \Idef{org27}Dipartimento di Fisica e Astronomia dell'Universit\`{a} and Sezione INFN, Bologna, Italy
\item \Idef{org28}Dipartimento di Fisica e Astronomia dell'Universit\`{a} and Sezione INFN, Catania, Italy
\item \Idef{org29}Dipartimento di Fisica e Astronomia dell'Universit\`{a} and Sezione INFN, Padova, Italy
\item \Idef{org30}Dipartimento di Fisica `E.R.~Caianiello' dell'Universit\`{a} and Gruppo Collegato INFN, Salerno, Italy
\item \Idef{org31}Dipartimento DISAT del Politecnico and Sezione INFN, Turin, Italy
\item \Idef{org32}Dipartimento di Scienze e Innovazione Tecnologica dell'Universit\`{a} del Piemonte Orientale and INFN Sezione di Torino, Alessandria, Italy
\item \Idef{org33}Dipartimento Interateneo di Fisica `M.~Merlin' and Sezione INFN, Bari, Italy
\item \Idef{org34}European Organization for Nuclear Research (CERN), Geneva, Switzerland
\item \Idef{org35}Faculty of Electrical Engineering, Mechanical Engineering and Naval Architecture, University of Split, Split, Croatia
\item \Idef{org36}Faculty of Engineering and Science, Western Norway University of Applied Sciences, Bergen, Norway
\item \Idef{org37}Faculty of Nuclear Sciences and Physical Engineering, Czech Technical University in Prague, Prague, Czech Republic
\item \Idef{org38}Faculty of Science, P.J.~\v{S}af\'{a}rik University, Ko\v{s}ice, Slovakia
\item \Idef{org39}Frankfurt Institute for Advanced Studies, Johann Wolfgang Goethe-Universit\"{a}t Frankfurt, Frankfurt, Germany
\item \Idef{org40}Gangneung-Wonju National University, Gangneung, Republic of Korea
\item \Idef{org41}Gauhati University, Department of Physics, Guwahati, India
\item \Idef{org42}Helmholtz-Institut f\"{u}r Strahlen- und Kernphysik, Rheinische Friedrich-Wilhelms-Universit\"{a}t Bonn, Bonn, Germany
\item \Idef{org43}Helsinki Institute of Physics (HIP), Helsinki, Finland
\item \Idef{org44}High Energy Physics Group,  Universidad Aut\'{o}noma de Puebla, Puebla, Mexico
\item \Idef{org45}Hiroshima University, Hiroshima, Japan
\item \Idef{org46}Hochschule Worms, Zentrum  f\"{u}r Technologietransfer und Telekommunikation (ZTT), Worms, Germany
\item \Idef{org47}Horia Hulubei National Institute of Physics and Nuclear Engineering, Bucharest, Romania
\item \Idef{org48}Indian Institute of Technology Bombay (IIT), Mumbai, India
\item \Idef{org49}Indian Institute of Technology Indore, Indore, India
\item \Idef{org50}Indonesian Institute of Sciences, Jakarta, Indonesia
\item \Idef{org51}INFN, Laboratori Nazionali di Frascati, Frascati, Italy
\item \Idef{org52}INFN, Sezione di Bari, Bari, Italy
\item \Idef{org53}INFN, Sezione di Bologna, Bologna, Italy
\item \Idef{org54}INFN, Sezione di Cagliari, Cagliari, Italy
\item \Idef{org55}INFN, Sezione di Catania, Catania, Italy
\item \Idef{org56}INFN, Sezione di Padova, Padova, Italy
\item \Idef{org57}INFN, Sezione di Roma, Rome, Italy
\item \Idef{org58}INFN, Sezione di Torino, Turin, Italy
\item \Idef{org59}INFN, Sezione di Trieste, Trieste, Italy
\item \Idef{org60}Inha University, Incheon, Republic of Korea
\item \Idef{org61}Institut de Physique Nucl\'{e}aire d'Orsay (IPNO), Institut National de Physique Nucl\'{e}aire et de Physique des Particules (IN2P3/CNRS), Universit\'{e} de Paris-Sud, Universit\'{e} Paris-Saclay, Orsay, France
\item \Idef{org62}Institute for Nuclear Research, Academy of Sciences, Moscow, Russia
\item \Idef{org63}Institute for Subatomic Physics, Utrecht University/Nikhef, Utrecht, Netherlands
\item \Idef{org64}Institute of Experimental Physics, Slovak Academy of Sciences, Ko\v{s}ice, Slovakia
\item \Idef{org65}Institute of Physics, Homi Bhabha National Institute, Bhubaneswar, India
\item \Idef{org66}Institute of Physics of the Czech Academy of Sciences, Prague, Czech Republic
\item \Idef{org67}Institute of Space Science (ISS), Bucharest, Romania
\item \Idef{org68}Institut f\"{u}r Kernphysik, Johann Wolfgang Goethe-Universit\"{a}t Frankfurt, Frankfurt, Germany
\item \Idef{org69}Instituto de Ciencias Nucleares, Universidad Nacional Aut\'{o}noma de M\'{e}xico, Mexico City, Mexico
\item \Idef{org70}Instituto de F\'{i}sica, Universidade Federal do Rio Grande do Sul (UFRGS), Porto Alegre, Brazil
\item \Idef{org71}Instituto de F\'{\i}sica, Universidad Nacional Aut\'{o}noma de M\'{e}xico, Mexico City, Mexico
\item \Idef{org72}iThemba LABS, National Research Foundation, Somerset West, South Africa
\item \Idef{org73}Johann-Wolfgang-Goethe Universit\"{a}t Frankfurt Institut f\"{u}r Informatik, Fachbereich Informatik und Mathematik, Frankfurt, Germany
\item \Idef{org74}Joint Institute for Nuclear Research (JINR), Dubna, Russia
\item \Idef{org75}Korea Institute of Science and Technology Information, Daejeon, Republic of Korea
\item \Idef{org76}KTO Karatay University, Konya, Turkey
\item \Idef{org77}Laboratoire de Physique Subatomique et de Cosmologie, Universit\'{e} Grenoble-Alpes, CNRS-IN2P3, Grenoble, France
\item \Idef{org78}Lawrence Berkeley National Laboratory, Berkeley, California, United States
\item \Idef{org79}Lund University Department of Physics, Division of Particle Physics, Lund, Sweden
\item \Idef{org80}Nagasaki Institute of Applied Science, Nagasaki, Japan
\item \Idef{org81}Nara Women{'}s University (NWU), Nara, Japan
\item \Idef{org82}National and Kapodistrian University of Athens, School of Science, Department of Physics , Athens, Greece
\item \Idef{org83}National Centre for Nuclear Research, Warsaw, Poland
\item \Idef{org84}National Institute of Science Education and Research, Homi Bhabha National Institute, Jatni, India
\item \Idef{org85}National Nuclear Research Center, Baku, Azerbaijan
\item \Idef{org86}National Research Centre Kurchatov Institute, Moscow, Russia
\item \Idef{org87}Niels Bohr Institute, University of Copenhagen, Copenhagen, Denmark
\item \Idef{org88}Nikhef, National institute for subatomic physics, Amsterdam, Netherlands
\item \Idef{org89}NRC Kurchatov Institute IHEP, Protvino, Russia
\item \Idef{org90}NRC «Kurchatov Institute»  - ITEP, Moscow, Russia
\item \Idef{org91}NRNU Moscow Engineering Physics Institute, Moscow, Russia
\item \Idef{org92}Nuclear Physics Group, STFC Daresbury Laboratory, Daresbury, United Kingdom
\item \Idef{org93}Nuclear Physics Institute of the Czech Academy of Sciences, \v{R}e\v{z} u Prahy, Czech Republic
\item \Idef{org94}Oak Ridge National Laboratory, Oak Ridge, Tennessee, United States
\item \Idef{org95}Ohio State University, Columbus, Ohio, United States
\item \Idef{org96}Petersburg Nuclear Physics Institute, Gatchina, Russia
\item \Idef{org97}Physics department, Faculty of science, University of Zagreb, Zagreb, Croatia
\item \Idef{org98}Physics Department, Panjab University, Chandigarh, India
\item \Idef{org99}Physics Department, University of Jammu, Jammu, India
\item \Idef{org100}Physics Department, University of Rajasthan, Jaipur, India
\item \Idef{org101}Physikalisches Institut, Eberhard-Karls-Universit\"{a}t T\"{u}bingen, T\"{u}bingen, Germany
\item \Idef{org102}Physikalisches Institut, Ruprecht-Karls-Universit\"{a}t Heidelberg, Heidelberg, Germany
\item \Idef{org103}Physik Department, Technische Universit\"{a}t M\"{u}nchen, Munich, Germany
\item \Idef{org104}Politecnico di Bari, Bari, Italy
\item \Idef{org105}Research Division and ExtreMe Matter Institute EMMI, GSI Helmholtzzentrum f\"ur Schwerionenforschung GmbH, Darmstadt, Germany
\item \Idef{org106}Rudjer Bo\v{s}kovi\'{c} Institute, Zagreb, Croatia
\item \Idef{org107}Russian Federal Nuclear Center (VNIIEF), Sarov, Russia
\item \Idef{org108}Saha Institute of Nuclear Physics, Homi Bhabha National Institute, Kolkata, India
\item \Idef{org109}School of Physics and Astronomy, University of Birmingham, Birmingham, United Kingdom
\item \Idef{org110}Secci\'{o}n F\'{\i}sica, Departamento de Ciencias, Pontificia Universidad Cat\'{o}lica del Per\'{u}, Lima, Peru
\item \Idef{org111}Shanghai Institute of Applied Physics, Shanghai, China
\item \Idef{org112}St. Petersburg State University, St. Petersburg, Russia
\item \Idef{org113}Stefan Meyer Institut f\"{u}r Subatomare Physik (SMI), Vienna, Austria
\item \Idef{org114}SUBATECH, IMT Atlantique, Universit\'{e} de Nantes, CNRS-IN2P3, Nantes, France
\item \Idef{org115}Suranaree University of Technology, Nakhon Ratchasima, Thailand
\item \Idef{org116}Technical University of Ko\v{s}ice, Ko\v{s}ice, Slovakia
\item \Idef{org117}Technische Universit\"{a}t M\"{u}nchen, Excellence Cluster 'Universe', Munich, Germany
\item \Idef{org118}The Henryk Niewodniczanski Institute of Nuclear Physics, Polish Academy of Sciences, Cracow, Poland
\item \Idef{org119}The University of Texas at Austin, Austin, Texas, United States
\item \Idef{org120}Universidad Aut\'{o}noma de Sinaloa, Culiac\'{a}n, Mexico
\item \Idef{org121}Universidade de S\~{a}o Paulo (USP), S\~{a}o Paulo, Brazil
\item \Idef{org122}Universidade Estadual de Campinas (UNICAMP), Campinas, Brazil
\item \Idef{org123}Universidade Federal do ABC, Santo Andre, Brazil
\item \Idef{org124}University of Cape Town, Cape Town, South Africa
\item \Idef{org125}University of Houston, Houston, Texas, United States
\item \Idef{org126}University of Jyv\"{a}skyl\"{a}, Jyv\"{a}skyl\"{a}, Finland
\item \Idef{org127}University of Liverpool, Liverpool, United Kingdom
\item \Idef{org128}University of Science and Techonology of China, Hefei, China
\item \Idef{org129}University of South-Eastern Norway, Tonsberg, Norway
\item \Idef{org130}University of Tennessee, Knoxville, Tennessee, United States
\item \Idef{org131}University of the Witwatersrand, Johannesburg, South Africa
\item \Idef{org132}University of Tokyo, Tokyo, Japan
\item \Idef{org133}University of Tsukuba, Tsukuba, Japan
\item \Idef{org134}Universit\'{e} Clermont Auvergne, CNRS/IN2P3, LPC, Clermont-Ferrand, France
\item \Idef{org135}Universit\'{e} de Lyon, Universit\'{e} Lyon 1, CNRS/IN2P3, IPN-Lyon, Villeurbanne, Lyon, France
\item \Idef{org136}Universit\'{e} de Strasbourg, CNRS, IPHC UMR 7178, F-67000 Strasbourg, France, Strasbourg, France
\item \Idef{org137}Universit\'{e} Paris-Saclay Centre d'Etudes de Saclay (CEA), IRFU, D\'{e}partment de Physique Nucl\'{e}aire (DPhN), Saclay, France
\item \Idef{org138}Universit\`{a} degli Studi di Foggia, Foggia, Italy
\item \Idef{org139}Universit\`{a} degli Studi di Pavia, Pavia, Italy
\item \Idef{org140}Universit\`{a} di Brescia, Brescia, Italy
\item \Idef{org141}Variable Energy Cyclotron Centre, Homi Bhabha National Institute, Kolkata, India
\item \Idef{org142}Warsaw University of Technology, Warsaw, Poland
\item \Idef{org143}Wayne State University, Detroit, Michigan, United States
\item \Idef{org144}Westf\"{a}lische Wilhelms-Universit\"{a}t M\"{u}nster, Institut f\"{u}r Kernphysik, M\"{u}nster, Germany
\item \Idef{org145}Wigner Research Centre for Physics, Hungarian Academy of Sciences, Budapest, Hungary
\item \Idef{org146}Yale University, New Haven, Connecticut, United States
\item \Idef{org147}Yonsei University, Seoul, Republic of Korea
\end{Authlist}
\endgroup

                                                                       
\end{document}